\documentclass[10pt]{article}

\usepackage[utf8]{inputenc} 
\usepackage[T1]{fontenc} 
\usepackage{microtype} 
\usepackage{times}
\usepackage{etoolbox}
\usepackage[pdftex]{graphicx}
\usepackage{color}
\usepackage[usenames,dvipsnames]{xcolor}
\usepackage{xspace}
\usepackage{geometry}
\usepackage[hyphens]{url}
\usepackage[bookmarks=true,hidelinks=false,linktocpage=false]{hyperref}
\usepackage{cite} 
\usepackage{hyphenat}
\usepackage[sort,numbers]{natbib}
\usepackage{algorithm}
\usepackage{algpseudocode}
\usepackage{amsthm}
\usepackage{amsmath}
\usepackage{booktabs}
\usepackage{balance}
\usepackage{enumerate}
\usepackage{lipsum}
\usepackage{titlesec}
\usepackage{wrapfig}
\usepackage{subfig}
\graphicspath{{./figs/}}

\newcommand{\etal}{\emph{et al.}\xspace}

\newenvironment{closeitemize}
{\begin{itemize}
    \setlength{\itemsep}{1pt}
    \setlength{\parskip}{0pt}
    \setlength{\parsep}{0pt}}
{\end{itemize}}

\newtheorem{problem}{Problem Formulation}

\usepackage{listings}

\lstdefinestyle{python}{
  language=Python,
  basicstyle=\ttm,
  otherkeywords={self},             
  keywordstyle=\ttb\color{deepblue},
  emph={MyClass,__init__},          
  emphstyle=\ttb\color{deepred},    
  stringstyle=\color{deepgreen},
  frame=tb,                         
  showstringspaces=false            %
}

\lstdefinestyle{bash}{
  language=bash,
  basicstyle=\small\sffamily,
  frame=tb,
  columns=fullflexible,
  backgroundcolor=\color{blue!5},
  linewidth=1.0\linewidth,
  xleftmargin=0.1\linewidth
}

\lstdefinestyle{java}{
  language=Java,
  basicstyle=\footnotesize\sffamily,
  numberstyle=\scriptsize,
  numbersep=5pt,
  tabsize=2,
  extendedchars=true,
  breaklines=true,
  commentstyle=\color{darkgreen}\textit,
  keywordstyle=\color{blue}\textbf,
  escapeinside={\%*}{*)},            
  columns=flexible
}

\newcommand{\bench}[1]{\textsf{\small #1}}
\newcommand{\code}[1]{\textsf{\small #1}}

\newcommand{\gem}{GEM\xspace}
\newcommand{\Gem}{\gem}
\newcommand{\capri}{Capri\xspace}
\newcommand{\Capri}{\capri}

\newtoggle{tech-report}
\toggletrue{tech-report}

\newtoggle{includeAcks}
\iftoggle{tech-report}{
  \toggletrue{includeAcks}
}{
  \togglefalse{includeAcks}
}

\begin{document}

\title{Capri: A Control System for Approximate Programs}

\author{
  Swarnendu Biswas\\
  \small University of Texas at Austin (USA)\\
  \small{sbiswas@ices.utexas.edu}
  \and
  Yan Pei\\
  \small University of Texas at Austin (USA)\\
  \small{ypei@cs.utexas.edu}
  \and
  Donald S. Fussell\\
  \small University of Texas at Austin (USA)\\
  \small{fussell@cs.utexas.edu}
  \and
  Keshav Pingali\\
  \small University of Texas at Austin (USA)\\
  \small pingali@cs.utexas.edu
}



\date{}
\maketitle

\begin{abstract}
  Approximate computing trades off accuracy of results for resources such as energy or computing time. There is a large
  and rapidly growing literature on approximate computing that has focused mostly on showing the benefits of approximation. However, we know relatively little about how to control approximation in a disciplined way.

  This document briefly describes our published work of controlling approximation for non-streaming programs that have a
  set of ``knobs'' that can be dialed up or down to control the level of approximation of different components in the
  program. The proposed system, \emph{Capri}, solves this control problem as a constrained optimization problem. Capri
  uses machine learning to learn cost and error models for the program, and uses these models to determine, for a
  desired level of approximation, knob settings that optimize metrics such as running time or energy usage. Experimental
  results with complex benchmarks from different problem domains demonstrate the effectiveness of this approach.

  This report outlines improvements and extensions to the existing Capri system to address its limitations, including a complete rewrite of the software, and discusses
  directions for follow up work. The document also includes instructions and guidelines for using the new Capri infrastructure.
\end{abstract}


\section{Introduction}
\label{sec:intro}

There is growing interest in \emph{approximate computing} as a way of reducing the energy and time required to execute
applications~\cite{baek-pldi-2010,ansel-cgo-2011, sidiroglou-douskos-fse-2011, sampson-pldi-2011,zhu-popl-2012}. In
conventional computing, programs are usually treated as implementations of mathematical functions, so there is a precise
output that must computed for a given input. In many problem domains, it is sufficient to produce some approximation of
this output; for example, when rendering a scene in graphics, it is acceptable to take computational short-cuts if human
beings cannot tell the difference in the rendered scene.

In this paper, we focus on a class of approximate programs that we call {\em tunable} approximate programs. Intuitively,
these programs have one or more \emph{knobs} or parameters that can be changed to vary the fidelity of the produced
output. These knobs might control the number of iterations performed by a loop~\cite{rinard-oopsla-2007,bottou-2010}, determine
the precision with which floating-point computations are performed~\cite{rubio-gonzalez-sc-2013,schkufza-pldi-2014}, or switch
between precise and approximate hardware~\cite{esmaeilzadeh-micro-2012}; for the purposes of this paper, the source of
approximation does not matter so long as the fidelity of the output is changed by adjusting the knobs.

There is now a fairly large literature on this subject, some of which is surveyed in Section~\ref{sec:relatedwork}. Most
of this work addresses what we call the \emph{forward problem} in this paper: they show that for some programs, particular
techniques such as skipping loop iterations or tasks, within limits, degrade output quality in an acceptable way while
reducing energy or running time. Other work has focused on type systems and static analyses to ensure that computational
short-cuts do not affect portions of the program that may be critical to correctness such as control-flow decisions or
memory management~\cite{sampson-pldi-2011,carbin-oopsla-2013,misailovic-oopsla-2014}.

However, exploiting approximation effectively requires the solution to what we call the \emph{inverse problem} in this
paper: given a program with knobs that control execution parameters like the number of the iterations executed by a loop
and a lower bound on output quality, how do we set the knobs optimally to minimize energy or running time? This is a
classical optimal control problem. What makes the problem particularly difficult is that for most programs, optimal
values of knob settings are very dependent on the values of inputs, as we show in Section~\ref{sec:problem}, so
auto-tuning, the standard parameter optimization technique used in computer systems, is not useful.

This paper describes our published work on solving the inverse problem for tunable approximate
programs~\cite{capri-asplos-2016}. Roughly speaking, given a permissible error for the output, we want to set the knobs
to minimize computational costs, such as running time or energy, while meeting the error constraint. The work describes
a solution to the proactive control problem for non-streaming programs that consist of components controlled by one or
more knobs and in which the error and cost behaviors are substantially different for different inputs. Our approach is
to treat the control problem as a \emph{constrained optimization} problem in which an objective function such as energy
is minimized, subject to constraints such as a lower bound on the acceptable output quality. The major challenge is that
this formulation requires us to know the objective and constraint functions, but in general these are complex functions
that we do not know and cannot write down in closed form. We deal with this by modeling these functions using machine
learning techniques. The resulting Capri control system~\cite{capri-asplos-2016}, which is an example of open-loop
control~\cite{astrom-2008}, is fairly successful in controlling approximation in a principled way in complex
applications from several domains including machine learning, image processing and graph analytics.


This paper extends the scope of our published work~\cite{capri-asplos-2016}, by first highlighting limitations of the
existing control system, such as, potential lack of scalability, and neglecting the prediction error from cost and error
models. We discuss follow up work to \capri that addresses these issues. A requisite for extending \capri is to
reimplement the system in a scalable and modular fashion. This paper discusses our new implementation in detail to
acquaint potential users with the internals of the \capri control system. We present approximation results with the new
\capri implementation.


\section{Related Work}
\label{sec:relatedwork}

\paragraph*{Approximation opportunities in software and hardware.}

Loop perforation~\cite{sidiroglou-douskos-fse-2011} explores skipping iterations during loop execution. Rinard explores
randomly discarding tasks in parallel applications~\cite{rinard-sc-2006}. Rinard~\cite{rinard-oopsla-2007} and Campanoni
\etal~\cite{campanoni-cgo-2015} explore relaxing synchronization in parallel applications. Karthik \etal
explore different algorithmic level approximation schemes on a video summarization algorithm~\cite{karthik-2015}.
Samadi \etal develop methods to recognize patterns in programs that provide approximation
opportunities~\cite{samadi-2014}. These techniques could be used to provide knobs automatically and thus complement our
work.

A distortion model using linear regression was used by Rinard to demonstrate the feasibility of their approximation
techniques~\cite{rinard-sc-2006}. The results in this paper (Section~\ref{sec:capri:results}) show that linear
regression is not useful for modeling quality and cost.

Researchers have proposed several hardware designs for exploiting approximate computing~\cite{palem-2005,
  esmaeilzadeh-asplos-2012, esmaeilzadeh-micro-2012, sampson-micro-2013, miguel-2014, shoushtari-2015}. Our techniques
can be useful in choosing how to most efficiently  map programs onto such hardware and thus increase the effectiveness
of such approaches.

\paragraph*{Reactive control of streaming applications.} In this problem, the system is presented with a stream of
inputs in which successive inputs are assumed to be correlated with each other, and results from processing one input
can be used to tune the computation for succeeding inputs. The Green System~\cite{baek-pldi-2010} periodically monitors
QoS values and recalibrates using heuristics whenever the QoS is lower than a specified level.
PowerDial~\cite{hoffmann-asplos-2011} leverages feedback control theory for recalibration. Argo~\cite{gadiolo-2015} is
an autotuning system for adapting application performance to changes in multicore resources. SAGE~\cite{sage-2013}
exploits this approach on GPU platforms. Fang \etal use simulated annealing to adjust the knob
settings~\cite{fang-taco-2014}. The problem considered in this paper is fundamentally different since it involves
proactive control of an application with a single input rather than reactive control for a stream of inputs. However,
the techniques described in this paper may be applicable to reactive control as well.

\paragraph*{Auto-tuning.}

Auto-tuning explores a space of exact implementations to optimize a cost metric like running time; in contrast, the
control problem defined in this paper deals with both error and cost dimensions. Several papers~\cite{ansel-cgo-2011,
  ding-pldi-2015} have extended the PetaBricks~\cite{ansel-pldi-2009} auto-tuning system to include an error bound. Ding
\etal group training inputs into clusters based on user-provided features, and auto-tuning is used to find
optimal knob settings for each cluster for given error bounds~\cite{ding-pldi-2015}. For a new input, optimal knob
settings for the same error bounds are determined by classifying the input into one of the clusters and using the
predetermined knob settings for that cluster. Auto-tuning is used by Precimonious~\cite{rubio-gonzalez-sc-2013} to lower
precision of floating point types to improve performance for a particular accuracy constraint.

The main difference between our approach and auto-tuning approaches is that our approach builds error and cost models
that can be used to control knobs for any error constraint presented during the online phase, without requiring
re-training. Since auto-tuning approaches do not build models, they do not have the ability to generalize their results
from the constraints they were trained for to other constraints. Note that the clustering-classification approach can be
combined with our approach by clustering the training inputs and building a different model for each cluster.

\paragraph*{Programming language support.}

EnerJ~\cite{sampson-pldi-2011} proposes a type system to separate exact and approximate data in the program.
Rely~\cite{carbin-oopsla-2013} uses static analysis techniques to quantify the errors in programs on approximate
hardware. Ringenburg \etal~\cite{ringenburg-asplos-2015} developed tools for debugging approximate programs. None of
these tools deal with controlling the tradeoff of error versus cost.

\paragraph*{Error guarantees.}

Zhu \etal formulated a randomized program transformation which trades off expected error versus performance as an
optimization problem~\cite{zhu-popl-2012}. However, their formulation assumes very small variations of errors across
inputs, an assumption violated in all of our complex real-world benchmark applications. They also assume the existence
of an \emph{a priori} error bound for each approximation in the program and that the error propagation is bounded by a
linear function. These assumptions make it hard to apply this approach to real-world applications. For example, we know
of no non-trivial error bounds for our benchmarks. Chisel~\cite{misailovic-oopsla-2014} extends
Rely~\cite{carbin-oopsla-2013} to use integer linear programming (ILP) to optimize the selection of instructions/data
executed/stored in approximate hardware. The ILP constraints are generated by static analysis, which propagates errors
through the program. While they consider input reliability, i.e. the probability that an input contains errors, they do
not deal with input sensitivity of the error function. Moreover, their error propagation method requires that the error
function be differentiable and their static analysis technique cannot deal with input-dependent loops, which are common
in our benchmarks and many other applications.

\emph{ApproxHadoop}~\cite{approxhadoop-2015} applies statistical sampling theory to Hadoop tasks for controlling input
sampling and task dropping. While statistical sampling theory gives nice error guarantees, the application of this
technique is restricted. Mahajan \etal~\cite{divya-2015} uses neural networks to predict whether to invoke approximate
accelerators or execute precise code for a quality constraint.

\paragraph*{Analytic properties of programs.}

Several techniques exist to verify whether a program is Lipschitz-continuous~\cite{chaudhuri-2012}. Smooth
interpretation~\cite{smooth-interpretation} can smooth out irregular features of a program. Given the input variability
exhibited in our applications, analytic properties usually provide very loose error bounds and are not helpful for
setting knobs.


\section{Problem Formulation}
\label{sec:problem}

We describe the formulation of the proactive control problem we use in this paper, justifying it by describing other
reasonable formulations and explaining why we do not use them. To keep notation simple, we consider a program that can
be controlled with two knobs $K_1$ and $K_2$ that take values from finite sets $\kappa_1$ and $\kappa_2$ respectively.
We write $K_1:\kappa_1$ and $K_2:\kappa_2$ to denote this, and use $k_1$ and $k_2$ to denote particular settings of
these knobs. The formulation generalizes to programs with an arbitrary number of knobs in an obvious way.

It is convenient to define the following functions.

\begin{closeitemize}
\item Output: In general, the output value of the tunable program
is a function of the input value $i$, and knob settings
$k_1$ and $k_2$. Let $f(i,k_1,k_2)$ be this function.

\item Error/quality degradation: Let $f_e(i,k_1,k_2)$ be the
magnitude of the output error or quality degradation
for input $i$ and knob settings $k_1$ and $k_2$.

\item Cost: Let $f_c(i,k_1,k_2)$ be the cost of computing the output
for input $i$ with knob settings $k_1$ and $k_2$. This can be the running time, energy or other execution metric to be optimized.
\end{closeitemize}

We formulate the control problem as an optimization problem in which the error is bounded for the particular input of
interest. This optimization problem is difficult to solve, so we formulate a different problem in which the expected
error over all inputs is less than the given error bound, with some probability. This gives the implementation
flexibility in finding low-cost solutions.

One way to formulate the control problem informally is the following: given an input value and a bound on the output
error, find knob settings that (i) meet the error bound and (ii) minimize the cost. This can be formulated as the
following constrained optimization problem.

\begin{problem}
\label{opt:form1}
Given:
\begin{closeitemize}
\item a program with knobs $K_1{:}\kappa_1$ and $K_2{:}\kappa_2$, and
\item a set of possible inputs I.
\end{closeitemize}

\noindent
For input $i {\in} I$ and error bound $\epsilon {>} 0$,
find $k_1 {\in} \kappa_1, k_2 {\in} \kappa_2$ such that
\begin{closeitemize}
\item $f_c(i,k_1,k_2)$ is minimized
\item $f_e(i,k_1,k_2) \leq \epsilon$
\end{closeitemize}
\end{problem}

In the literature, the constraint $f_e(i,k_1,k_2) \leq \epsilon$ is said to define the {\em feasible region}, and
values of $(k_1,k_2)$ that satisfy this constraint for a given input are said to lie within the feasible region for that
input. The function $f_c(i,k_1,k_2)$ is the {\em objective function}, and a solution to the optimization problem is a
point that lies within the feasible region and minimizes the objective function.

\begin{figure}
  \minipage{0.49\textwidth}
  \centering
  \includegraphics[scale=0.52]{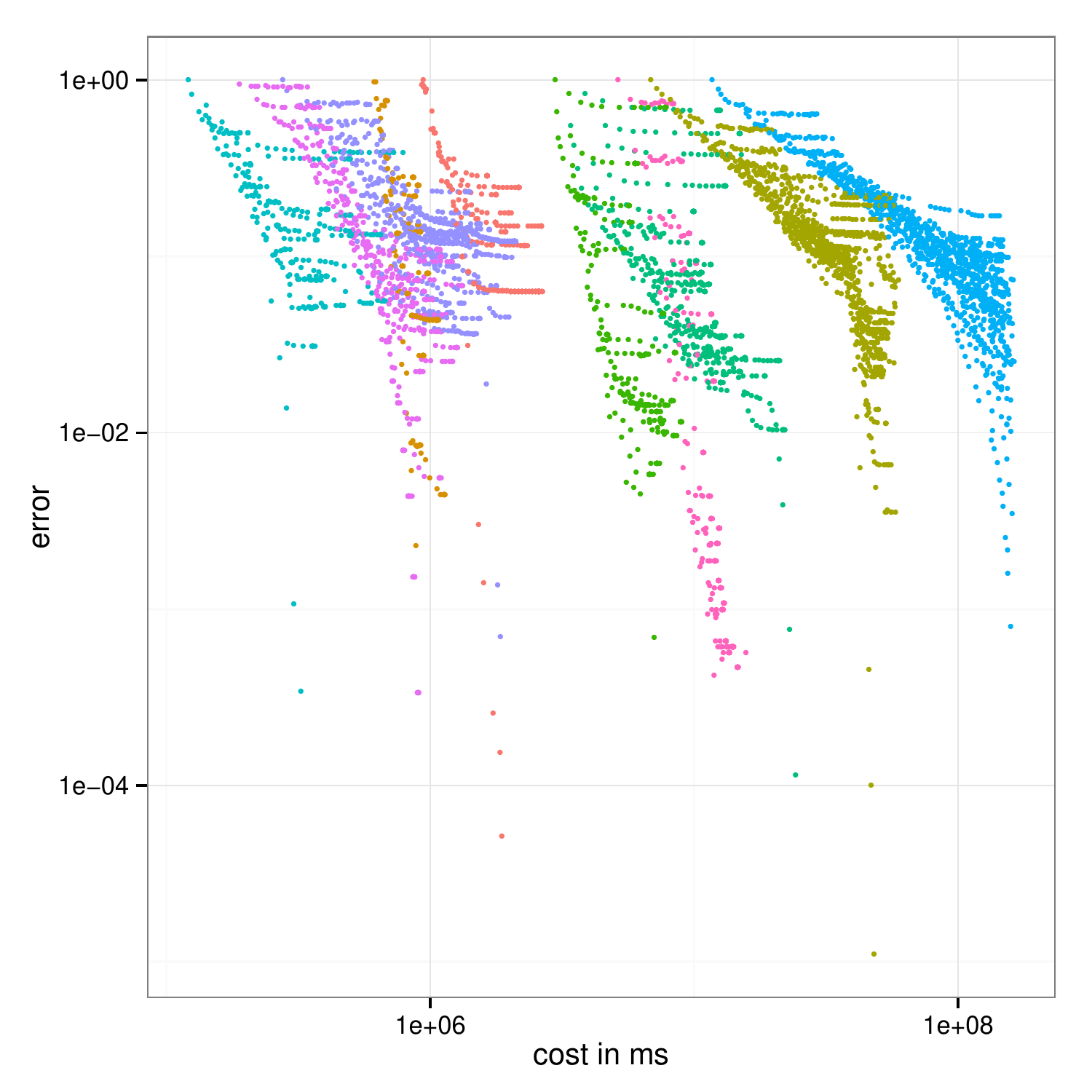}
  \caption{Cost vs. error for GEM. Each dot represents one knob setting for one input. Different colors
    represent different inputs.}
  \label{fig:costvsError}
  \endminipage
  \hfill
  \minipage{0.49\textwidth}
  \centering
  \includegraphics[scale=0.52]{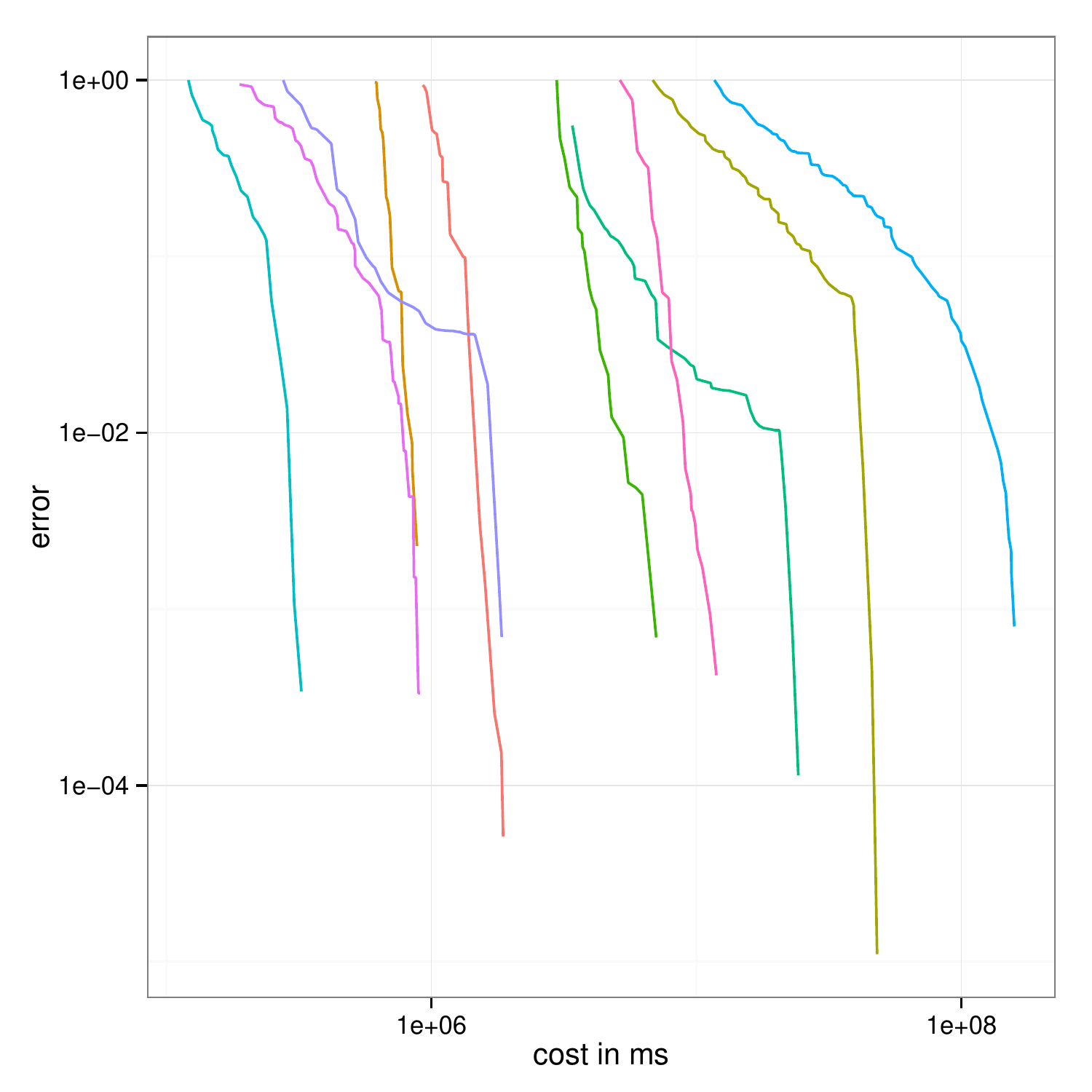}
  \caption{Pareto-optimal curves for GEM benchmark. Different lines represent different inputs.}
  \hspace{5pt}
  \label{fig:pareto}
  \endminipage\hfill
\end{figure}

For most tunable programs, this is a very complex optimization problem since the Pareto-optimal knob settings vary
greatly for different inputs~\cite{capri-asplos-2016}. To get a sense of this complexity, consider the \gem benchmark, a
graph partitioner for social network graphs~\cite{whang-icdm-2012} studied in more detail in
Sections~\ref{sec:capri:results} and~\ref{sec:applications}. Figure~\ref{fig:costvsError} shows the results of running
\gem with a variety of inputs and different knob settings, and measuring the cost (running time of the program) and
error of the output of the resulting programs. In this figure, each point represents the cost and error for a single
input graph and knob settings combination; points that correspond to the same input graph are colored identically. It
can be seen that even for a single input graph, there are many knob combinations that produce the same output error, and
that these combinations have widely different costs.

For a given input graph and output error, we are interested in minimizing cost, so only the leftmost point for each such
combination is of interest. Figure~\ref{fig:pareto} shows these \emph{Pareto-optimal} points for each input graph. Since
these Pareto-optimal curves are very different for different inputs, it is difficult to produce the Pareto-optimal knob
settings for a given input and output error without exploring much of the space of knob settings for a given input,
which is intractable for non-trivial systems.

One way to simplify the control problem is to require only that the \emph{expected} output error over all inputs be less
than some specified bound $\epsilon$. Since some inputs may be more likely to be presented to the system than others,
each input can be associated with a probability that is the likelihood that input is presented to the system. This lets
us give more weight to more likely inputs, as is done in Valiant's probably approximately correct (PAC) theory of
machine learning~\cite{valiant-1984}. Since the cost function is still a function of the actual input, knob settings for
a given value of $\epsilon$ will be different in general for different inputs, but the output error will be within the
given error bounds only in an average sense. In our approach, we consider a variation of this optimization problem,
inspired by Valiant's work~\cite{valiant-1984}, in which we are also given a probability $\pi$ with which the error
bound must be met. Intuitively, values of $\pi$ less than 1 give the control system a degree of slack in meeting the
error constraint, permitting the system to find lower cost solutions. This control problem can be formulated as an
optimization problem as follows.
\pagebreak
\begin{problem}
  \label{opt:final}
  Given:
  \begin{closeitemize}
  \item a program with knobs $K_1:\kappa_1$ and $K_2:\kappa_2$,
  \item a set of possible inputs $I$, and
  \item a probability function $p$ such that for any $i \in I$, $p(i)$ is the probability of getting input $i$.
  \end{closeitemize}

  \noindent
  For an input $i \in I$, error bound $\epsilon > 0$, and a probability \mbox{$1 \geq \pi > 0$ } with which this error
  bound must be met, find $k_1 \in \kappa_1$, $k_2 \in \kappa_2$ such that
  \begin{closeitemize}
  \item $f_c(i,k_1,k_2)$ is minimized
  \item $\sum\limits_{(j \in I)\wedge (f_e(j,k_1,k_2) \leq \epsilon)} p(j)\ \ \ \geq \pi$
  \end{closeitemize}
\end{problem}

If the term $\footnotesize \sum\limits_{(j \in I)\wedge (f_e(j,k_1,k_2) \leq \epsilon)} p(j)$ (denoted by
$P_{e}(\epsilon, k_{1}, k_{2})$) is greater or equal to $\pi$, then $(k_{1}, k_{2})$ is in the feasible region for error
bound $\epsilon$. For future reference, we call this the \emph{fitness} of knob setting $(k_{1}, k_{2})$ for error
$\epsilon$; intuitively, the greater the fitness of a knob setting, the more likely it is that it satisfies the error
bound for the given ensemble of inputs. In the rest of the paper, we refer to Problem Formulation~\ref{opt:final} as
the ``control problem.''


\section{\Capri: Proactive Control for Approximate Programs}
\label{sec:capri}

\begin{figure}
  \centering
   \includegraphics[scale=0.4]{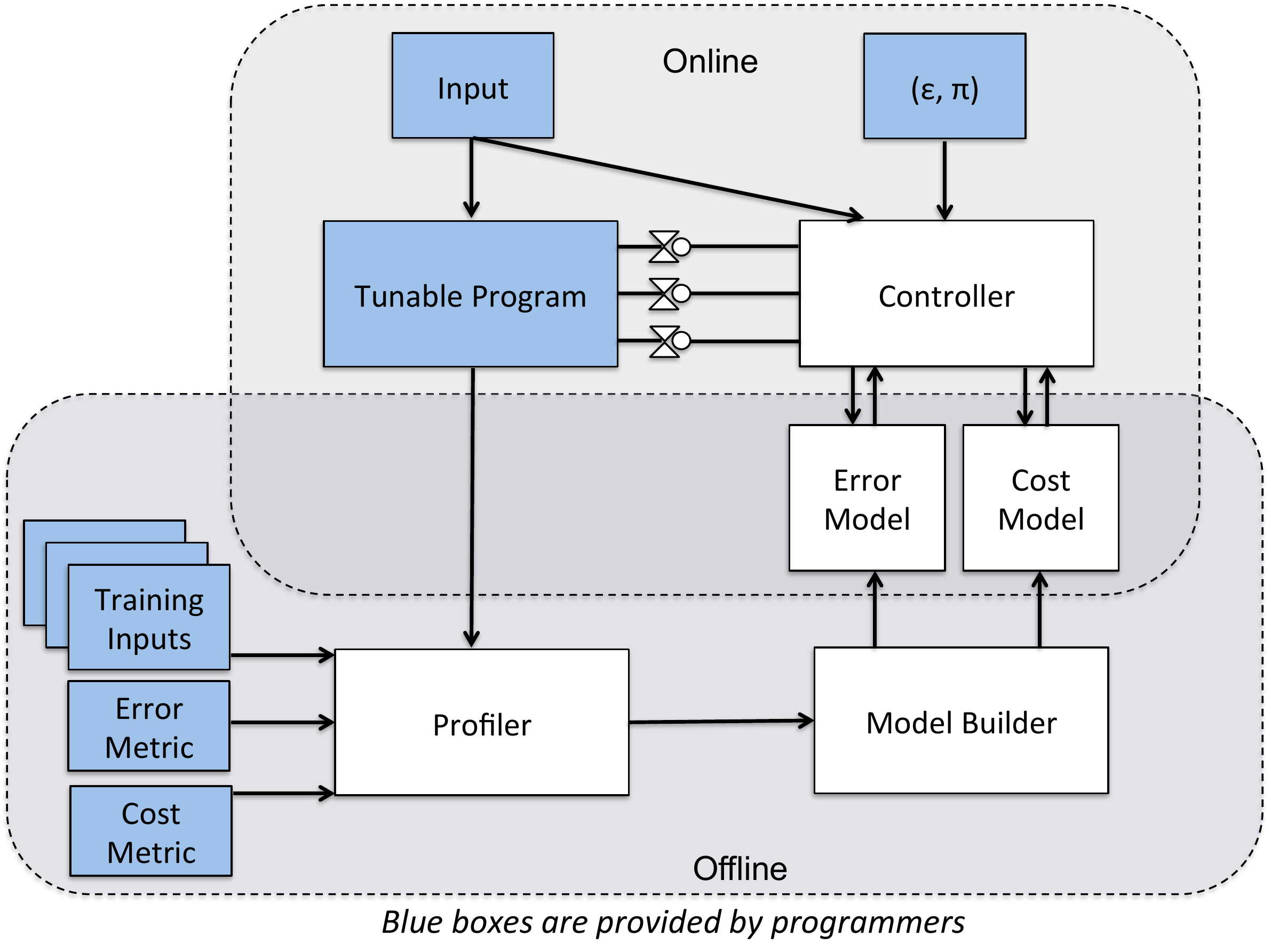}
  \caption{Overview of the Capri control system}
  \label{fig:architecture}
\end{figure}

For the complex applications we are interested in, the error function $f_e(i, k_{1}, k_{2})$ and the cost function
$f_c(i, k_{1}, k_{2})$ are non-linear functions of the inputs, and it is difficult if not impossible to derive
closed-form expressions for them. Therefore, we use machine learning techniques to build proxies for these functions
offline, using a suitable collection of training inputs. Figure~\ref{fig:architecture} is an overview of the control
system, which we call \emph{Capri}. For a given program, the system must be provided with a set of training inputs, and
metrics for the error/quality of the output and the cost. The offline portion of the system runs the program on these
inputs using a variety of knob settings, and learns the functions $f_e$ and $f_c$. These models are inputs to the
controller in the online portion of the system; given an input and values of $\epsilon$ and $\pi$, the controller solves
the control problem to estimate optimal knob settings. In the following, we describe the important modules of the \capri
control system.

\subsection{Error Model}
\label{sec:error-model}

The error model is a proxy for the fitness function $P_{e}(\epsilon, k_{1}, k_{2})$ and is used to determine whether a
knob setting is in the feasible region. Intuitively, a knob setting is in the feasible region if the inputs for which
the error is between $0$ and $\epsilon$ have a combined probability mass greater than or equal to $\pi$. We use Bayesian
networks~\cite{neapolitan} to determine this. A Bayesian network is a directed acyclic graph (DAG) in which each node
represents a random variable in the model and each edge represents the dependence relationship between the variables
corresponding to the nodes of its end points.

There are several ways to model the error probability distribution using a Bayesian network. We use a simple model in
which each of the knobs and the error is modeled as a random variable and the output error depends on all of the knobs.
The disadvantage of this simple model is that the size of the table for the output error is exponential in the number of
knobs (see Section~\ref{subsec:scale-large-tunable}); however, it works well for the applications we have investigated. Our
system allows new models for error to be plugged in easily into the overall framework (Section~\ref{subsec:scale-large-tunable}).

\subsection{Cost Model}
\label{sec:cost-model}

The cost model is the proxy for $f_c(i,k_{1},k_{2})$. We model both the running time and total energy. For most
algorithms, the running time can vary substantially for different inputs; after all, even for simple algorithms like
matrix multiplication, the running time is a function of the input size. For complex irregular algorithms like the ones
considered in this proposal, running time will depend not only the input size but also on other features of the input.
For example, the running time of a graph clustering algorithm is affected by the number of vertices and edges in the
graph as well as the number of clusters. Therefore, the running time is usually a complex function of input {\em
  features} and knob settings. Our system currently requires the user to specify what these features are.

We use M5~\cite{quinlan-92}, which builds tree-based models, to model the cost function $f_c$. Input features and knob
settings define a multidimensional space; the tree model divides this space into a set of subspaces, and constructs a
linear model in each subspace. The division into sub-spaces is done automatically by M5, which is a major advantage of
using this system. Intuitively, this model can approximate cost well because the running time does not usually exhibit
sharp discontinuities with respect to knob settings.

\subsection{Controller}
\label{sec:controller}

The control algorithm must search the space of knob settings to find optimal knob settings, using the error and cost
models as proxies for $f_e$ and $f_c$ respectively. Our system is implemented so that new search strategies can be
incorporated seamlessly. This lets us evaluate model accuracy separately from search accuracy.

We evaluated two search algorithms: \emph{exhaustive search} and \emph{Precimonious
  search}~\cite{rubio-gonzalez-sc-2013}. If the error and cost models are not expensive to evaluate and each knob has a
finite number of settings, we can use exhaustive search. We sweep over the entire space of knob settings, and for each
knob setting, use the error model to determine if that knob setting is in the feasible region. The cost model is then
used to find a minimal cost point in the feasible region. In a large search space, heuristics-based search is an
effective way to trade-off search cost for quality of the result. Precimonious search, which is based on the
delta-debugging algorithm~\cite{hildebrandt-issta-2000}, is one such strategy. The algorithm starts with all knobs set
to the highest values, and attempts to lower these settings iteratively. Precimonious can quickly prune the search space
but the solution it finds may be a local minimum. Other search strategies can be implemented easily within Capri.

\subsection{Results}
\label{sec:capri:results}

We evaluate the control system on the following five complex applications: (i) \gem, a graph partitioner for social
networks~\cite{whang-icdm-2012}, (ii) Ferret~\cite{bienia-2011}, a content-similarity based image search engine, (iii)
ApproxBullet~\cite{otaduy-sgp-2003}, a 3D physics game engine, (iv) SGDSVM~\cite{bottou-2010}, a library for support
vector machines, and (v) OpenOrd~\cite{openord-2011}, a library for two-dimensional graph layouts. We modified the code
for these applications to permit control of approximations. These applications provide between two and five knobs that
allow tuning tradeoffs between a user-specified quality metric in each case and the execution time or energy
consumption. In addition, we did a blind test of the system using an unmodified radar processing
application~\cite{hoffmann-tpds-2012}, written by Hank Hoffmann at the University of Chicago.

\paragraph*{Evaluation of the cost and error models.}

For each benchmark, we collected a set of inputs. To evaluate the error and the cost models, the inputs were randomly
partitioned into training and testing subsets. We evaluated our control system for $\epsilon$ ranging from $0.0$ to
$1.0$ and $\pi$ ranging from $0.1$ to $1.0$. Training is done offline, so training time is not as important as the
accuracy of the cost and error models. Training time obviously increases with the number of training inputs, but even
for Ferret, which has the largest training set, it takes only $0.927$ seconds to train the error model and $133.366$
seconds (about 2 minutes) to train the cost model (this does not take into account the time to run the application
programs).

\begin{figure}
  \centering
  \includegraphics[scale=0.8]{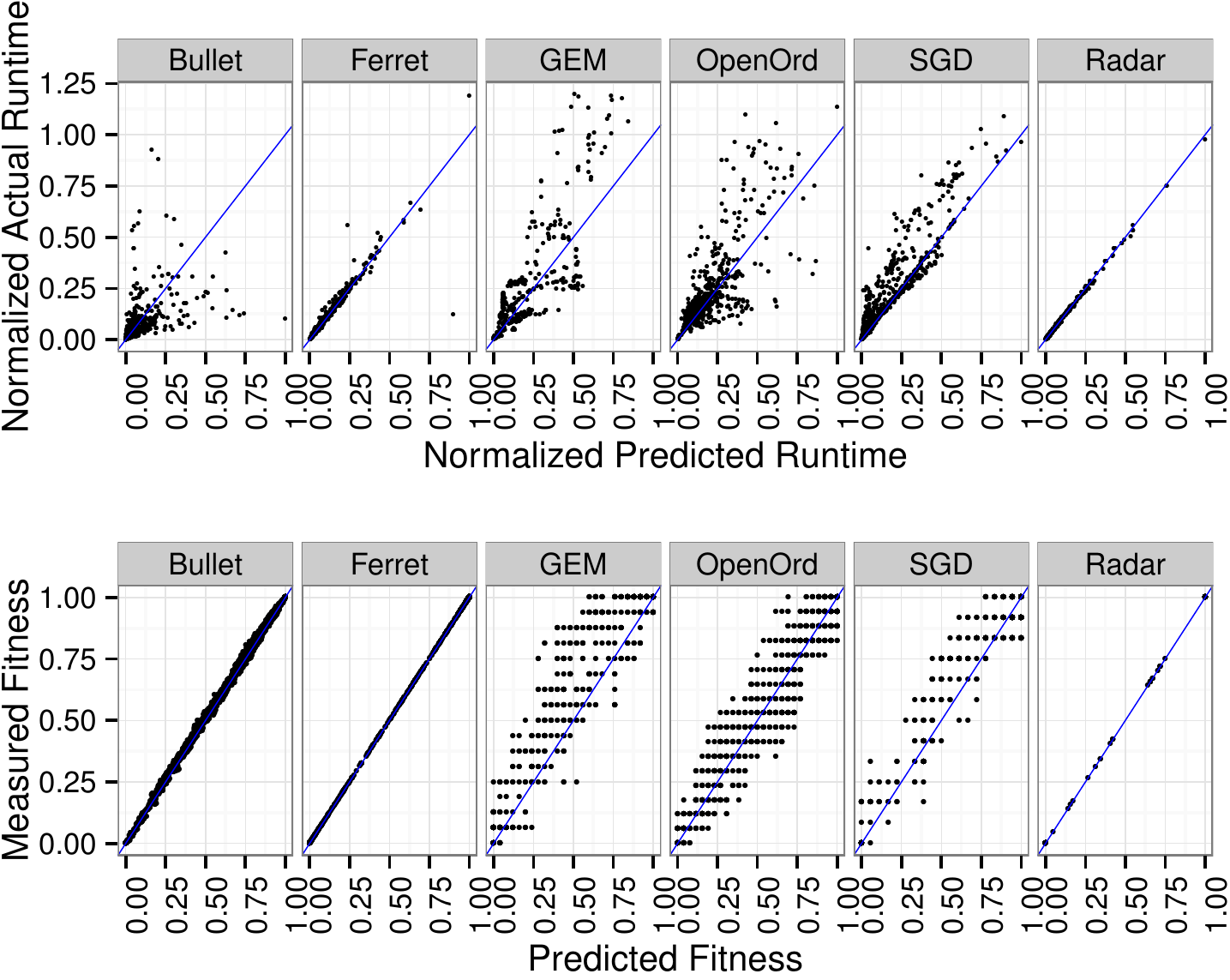}
  \caption{Accuracy of cost and error models}
  \label{fig:modelAccuracy}
\end{figure}

Evaluating the accuracy of the cost model for a given application is straightforward: we sweep the space of test inputs and knob settings, and for each point in this space, we compare the running time predicted by the cost model with the actual execution time. The top charts in Figure~\ref{fig:modelAccuracy} show the results for the applications in our
test suite. In each graph, the x-axis is the predicted running time and the y-axis is the measured running time. If the cost model is perfect, all points should lie on the $y{=}x$ line. Figure~\ref{fig:modelAccuracy} shows that this is more or less true for Ferret and Radar. For \gem and SGD, the predicted time is usually less than the actual execution time, and for Bullet and OpenORD, the over-predictions and under-predictions are more or less evenly distributed. Radar implements a regular algorithm in which running time depends on the size of the input. In contrast, \gem and OpenORD implement complex graph algorithms, so they are more irregular in their behavior.

Estimating the accuracy of the error model has to be more indirect since the model does not make error predictions for individual inputs but only for an ensemble of inputs $I$. The error model is a proxy for the fitness function
$P_e(\epsilon,k_1,k_2)$. This proxy is constructed during the training phase by letting $I$ be the set of training inputs. One way to evaluate the accuracy of this proxy is to construct another proxy by letting $I$ be the set of test inputs. If the model is accurate, these two proxy functions, which we call the predicted fitness and measured fitness, will be equal.

The bottom charts of Figure~\ref{fig:modelAccuracy} show the results of this experiment. The x and y axes in each graph
are the predicted and measured fitness respectively. We sweep over the space of (discretized) error values $\epsilon$
and knob settings, and for each point in this space, we evaluate the two proxy functions and plot the point in the
graph. We see that the error model is very accurate for Bullet, Ferret and Radar, and less so for the other three
benchmarks. For \gem and SGD, most of the points lie above the $y{=}x$ line, which means that the predicted fitness is
usually less than the actual fitness. Therefore, the feasible region determined by using the model may be smaller than
the actual feasible region.

It is important to note that since the error and cost models are used only to \emph{rank} knob settings in the feasible region,
more accurate models do not necessarily give better solutions to the control problem.

\paragraph*{Optimizing run-time performance.}

\begin{table}
  \footnotesize
  \centering
  \begin{tabular}
    {l|@{\hspace{0.05cm}}r@{\hspace{0.05cm}}r@{\hspace{0.05cm}}r@{\hspace{0.05cm}}r@{\hspace{0.05cm}}r@{\hspace{0.05cm}}r|@{\hspace{0.05cm}}r@{\hspace{0.05cm}}r@{\hspace{0.05cm}}r@{\hspace{0.05cm}}r@{\hspace{0.05cm}}r@{\hspace{0.05cm}}r|@{\hspace{0.05cm}}r@{\hspace{0.05cm}}r@{\hspace{0.05cm}}r@{\hspace{0.05cm}}r@{\hspace{0.05cm}}r@{\hspace{0.05cm}}r|@{\hspace{0.05cm}}r@{\hspace{0.05cm}}r@{\hspace{0.05cm}}r@{\hspace{0.05cm}}r@{\hspace{0.05cm}}r@{\hspace{0.05cm}}r|@{\hspace{0.05cm}}r@{\hspace{0.05cm}}r@{\hspace{0.05cm}}r@{\hspace{0.05cm}}r@{\hspace{0.05cm}}r@{\hspace{0.05cm}}r}  \hline
    & \multicolumn{6}{c}{Bullet} & \multicolumn{6}{c}{Ferret} & \multicolumn{6}{c}{GEM} & \multicolumn{6}{c}{OpenOrd} & \multicolumn{6}{c}{SGD} \\ \hline
    $\pi\parallel\epsilon$&0.0 & 0.1 & 0.2 & 0.3 & 0.4 & 0.5 & 0.0 & 0.1 & 0.2 & 0.3 & 0.4 & 0.5 & 0.0 & 0.1 & 0.2 & 0.3 & 0.4 & 0.5 & 0.0 & 0.1 & 0.2 & 0.3 & 0.4 & 0.5 & 0.0 & 0.1 & 0.2 & 0.3 & 0.4 & 0.5\\\hline
    1.0 & 1.0 & 1.0 & 1.0 & 1.0 & 1.0 & 1.0 & 1.0 & 1.0 & 1.0 & 1.0 & 1.1 & 1.1 & NA & NA & 1.3 & 1.4 & 1.6 & 1.9 & NA & 2.0 & 2.4 & 6.3 & 6.3 & 5.9 & NA & 2.5 & 5.6 & 14.1 & 31.3 & 77.4\\
    0.9 & 1.0 & 1.0 & 1.0 & 1.0 & 1.0 & 1.0 & 1.1 & 1.4 & 1.4 & 1.6 & 1.7 & 1.9 & NA & 1.1 & 1.5 & 1.9 & 2.2 & 2.5 & NA & 2.9 & 6.5 & 6.0 & 5.9 & 8.4 & NA & 11.8 & 30.3 & 43.0 & 56.3 & 94.9\\
    0.8 & 1.0 & 1.0 & 1.0 & 1.0 & 1.0 & 24.3 & 1.1 & 1.4 & 1.6 & 1.6 & 1.9 & 1.9 & NA & 1.2 & 1.7 & 2.1 & 2.4 & 2.6 & NA & 5.2 & 6.4 & 8.7 & 8.7 & 8.5 & NA & 39.7 & 52.5 & 103.7 & 165.0 & 184.0\\
    0.7 & 1.0 & 1.0 & 1.0 & 1.0 & 9.0 & 96.6 & 1.1 & 1.4 & 1.6 & 1.7 & 1.9 & 2.0 & NA & 1.3 & 1.7 & 2.1 & 2.5 & 2.7 & NA & 6.3 & 6.1 & 8.5 & 8.8 & 8.5 & NA & 73.3 & 101.4 & 139.9 & 176.1 & 271.7\\
    0.6 & 1.0 & 1.0 & 1.0 & 1.7 & 96.6 & 141.3 & 1.2 & 1.4 & 1.6 & 1.7 & 1.9 & 2.0 & NA & 1.4 & 2.0 & 2.2 & 2.6 & 2.7 & NA & 6.0 & 8.5 & 8.7 & 8.5 & 8.5 & 1.0 & 97.5 & 136.7 & 207.0 & 302.0 & 395.3\\
    0.5 & 1.0 & 1.0 & 1.0 & 39.9 & 115.4 & 204.6 & 1.2 & 1.4 & 1.6 & 1.7 & 2.0 & 2.0 & NA & 1.7 & 2.3 & 2.5 & 2.7 & 3.0 & NA & 8.5 & 8.5 & 8.6 & 8.4 & 8.4 & 1.0 & 104.6 & 161.1 & 259.1 & 302.0 & 418.4\\
    \hline

  \end{tabular}
  \caption{Speedups of the tuned programs for a subset of constraint space. }
  \label{speedups-tuned}
\end{table}


Speedup is defined as ratio of the running time at a particular knob setting to the running time with the knobs set for
maximum quality. Table~\ref{speedups-tuned} shows speedups for each application for $\epsilon$ values between 0 and 0.5
and $\pi$ values between 0.5 and 1.0. 
Each entry gives the average speedup over all test inputs for the knob settings found by the control algorithm based on
exhaustive search, given $(\epsilon,\pi)$ constraints in the intervals specified by the row and column indices.

Speedups depend on the application and the $(\epsilon, \pi)$ constraints. For each application, the top-left corner of
the constraint space is the ``hard'' region since the error must be low with high probability. The knob settings must be
at or close to maximum, and speedup will be limited.
Table entries marked ``NA'' show where the control system was unable to find any feasible solution for these hard constraints. 
In contrast, the bottom-right corner of the constraint space is the ``easier'' region, so one would expect higher speedups. This is seen in all benchmarks.
Overall, we see that controlling the knobs in these applications can yield significant speedups in running time.

\paragraph*{Effectiveness in finding optimal knob settings.}
\label{sec:optimize:results:optimal-knob}

While Table~\ref{speedups-tuned} shows speedups obtained from the knob settings found by the control algorithm in
different regions of the constraint space, it does not show how well these constraints were actually met. To provide
context, we have evaluated this both for our method and for a similar method using linear regression to model both error
and running time (linear regression can be seen as the simplest non-trivial model one can build for these values). For
each ($\epsilon,\pi$) combination, we evaluated the quality of the achieved control.

Overall, the control system using the Bayes model for error and the M5 model for cost performs quite well for all inputs
and regions of the constraint space: for most points, it finds solutions and the cost difference from \emph{the
  oracle}'s solution is within 40\%. The only noticeable problem is in SGD. A closer study showed that the feasible
region found by the Bayes error model is smaller than it should be and did not contain some low-cost points found by the
oracle control. This can be attributed to the fact that the predicted fitness function for SGD is somewhat conservative,
as seen in Figure~\ref{fig:modelAccuracy}.

In contrast, the control system based on linear regression performs quite poorly. No solutions are found in most parts
of the space, and even when solutions are found, the cost of the solutions is very sub-optimal.

\paragraph*{Performance of the Radar processing application.}

We also performed a blind test of the system using a radar processing application~\cite{hoffmann-tpds-2012}. Unlike the
five applications described above, this code was already instrumented with knobs, so we used it out of the box as a
blind test for our system. Using our machine-learning-based control scheme, we were able to obtain speedups over a base
fixed system configuration comparable to those obtained by hand tuning. In contrast, models using linear regression were
unable to find solutions in most of the constraint space.

\paragraph*{Optimizing energy consumption.}

We note that a major advantage of our approach is that it can be used to optimize not just running time but \emph{any
  metric for which a reasonable cost model can be constructed}. 
In this section, we show the results of applying the system to optimizing energy consumption for the same benchmarks. We
measured energy on a Intel Xeon E5-2630 CPU with 16 GB of memory. We used the Intel RAPL (Running Average Power Limit)
interface and PAPI to measure the energy consumption. This machine does not support DRAM counters, so what is being
measured is the CPU package energy consumption.

Table~\ref{power-savings} shows the power savings obtained for our benchmarks for $\epsilon$ values between 0 and 0.5
and $\pi$ values between 0.5 and 1.0. Each entry gives the average power savings over all test inputs for the knob
settings found by our control algorithm given $(\epsilon,\pi)$ constraints in the intervals specified by the row and
column indices. As expected, savings are greater when the constraints are looser.

\begin{table}
\scriptsize
\centering
\begin{tabular}
{l|@{\hspace{0.05cm}}r@{\hspace{0.05cm}}r@{\hspace{0.05cm}}r@{\hspace{0.05cm}}r@{\hspace{0.05cm}}r@{\hspace{0.05cm}}r|@{\hspace{0.05cm}}r@{\hspace{0.05cm}}r@{\hspace{0.05cm}}r@{\hspace{0.05cm}}r@{\hspace{0.05cm}}r@{\hspace{0.05cm}}r|@{\hspace{0.05cm}}r@{\hspace{0.05cm}}r@{\hspace{0.05cm}}r@{\hspace{0.05cm}}r@{\hspace{0.05cm}}r@{\hspace{0.05cm}}r|@{\hspace{0.05cm}}r@{\hspace{0.05cm}}r@{\hspace{0.05cm}}r@{\hspace{0.05cm}}r@{\hspace{0.05cm}}r@{\hspace{0.05cm}}r|@{\hspace{0.05cm}}r@{\hspace{0.05cm}}r@{\hspace{0.05cm}}r@{\hspace{0.05cm}}r@{\hspace{0.05cm}}r@{\hspace{0.05cm}}r|@{\hspace{0.05cm}}r@{\hspace{0.05cm}}r@{\hspace{0.05cm}}r@{\hspace{0.05cm}}r@{\hspace{0.05cm}}r@{\hspace{0.05cm}}r}  \hline
 & \multicolumn{6}{c}{Bullet} & \multicolumn{6}{c}{Ferret} & \multicolumn{6}{c}{GEM} & \multicolumn{6}{c}{OpenOrd} & \multicolumn{6}{c}{SGD} & \multicolumn{6}{c}{Radar} \\ \hline
$\pi\parallel\epsilon$&0.0 & 0.1 & 0.2 & 0.3 & 0.4 & 0.5 & 0.0 & 0.1 & 0.2 & 0.3 & 0.4 & 0.5 & 0.0 & 0.1 & 0.2 & 0.3 & 0.4 & 0.5 & 0.0 & 0.1 & 0.2 & 0.3 & 0.4 & 0.5 & 0.0 & 0.1 & 0.2 & 0.3 & 0.4 & 0.5 & 0.0 & 0.1 & 0.2 & 0.3 & 0.4 & 0.5\\\hline
1.0 & 1.0 & 1.0 & 1.0 & 1.0 & 1.0 & 1.0 & NA & 1.0 & 1.0 & 1.0 & 1.1 & 1.1 & NA & NA & 1.3 & 1.5 & 1.7 & 1.9 & NA & 2.4 & 6.1 & 7.2 & 7.2 & 8.9 & NA & 21.6 & 59.5 & 83.3 & 108.3 & 107.3 & 1.0 & 1.0 & 1.0 & 1.1 & 1.1 & 1.1\\
0.9 & 1.0 & 1.0 & 1.1 & 1.0 & 1.0 & 1.0 & 1.0 & 1.5 & 1.7 & 1.8 & 1.8 & 1.8 & NA & NA & 1.7 & 2.0 & 2.1 & 2.3 & NA & 6.0 & 7.1 & 7.2 & 8.9 & 8.9 & NA & 51.0 & 98.0 & 149.2 & 168.7 & 262.7 & 1.0 & 1.0 & 1.0 & 1.1 & 1.1 & 1.1\\
0.8 & 1.0 & 1.1 & 1.0 & 1.0 & 1.0 & 1.0 & 1.1 & 1.6 & 1.8 & 1.8 & 1.8 & 1.8 & NA & 1.1 & 1.8 & 2.1 & 2.3 & 2.5 & NA & 6.0 & 7.2 & 8.9 & 8.9 & 8.9 & NA & 91.0 & 192.5 & 266.0 & 265.0 & 319.0 & 1.0 & 1.0 & 1.0 & 1.1 & 1.1 & 1.1\\
0.7 & 1.0 & 1.0 & 1.0 & 1.0 & 1.0 & 1.0 & 1.1 & 1.7 & 1.8 & 1.8 & 1.8 & 1.8 & NA & 1.2 & 1.8 & 2.3 & 2.5 & 2.5 & NA & 7.1 & 7.2 & 8.9 & 8.9 & 8.9 & NA & 112.7 & 193.6 & 265.0 & 338.2 & 319.0 & 1.0 & 1.0 & 1.0 & 1.1 & 1.1 & 1.1\\
0.6 & 1.0 & 1.0 & 1.0 & 1.0 & 1.0 & 1.2 & 1.4 & 1.8 & 1.8 & 1.8 & 1.8 & 1.8 & NA & 1.5 & 2.1 & 2.3 & 2.5 & 2.8 & NA & 7.2 & 8.9 & 8.9 & 8.9 & 8.9 & 1.0 & 110.2 & 193.6 & 345.1 & 341.8 & 410.2 & 1.0 & 1.0 & 1.0 & 1.3 & 1.3 & 1.3\\
0.5 & 1.0 & 1.0 & 1.0 & 1.0 & 1.2 & 1.3 & 1.4 & 1.8 & 1.8 & 1.8 & 1.8 & 1.8 & NA & 1.7 & 2.3 & 2.5 & 2.8 & 2.8 & NA & 8.9 & 8.9 & 8.9 & 8.9 & 8.9 & 1.0 & 129.9 & 254.2 & 345.1 & 420.2 & 410.2 & 1.0 & 1.0 & 1.0 & 1.3 & 1.3 & 1.3\\
\hline
\end{tabular}
\caption{Energy savings of the tuned programs for a subset of constraint space. }
\label{power-savings}
\end{table}


\section{Extending Capri}
\label{sec:extensions}

The Capri system is an example of an \emph{open-loop control system}, which uses a model of the system (in our case, the
tunable program) to determine optimal knob settings \emph{before} the application is executed. However, the \capri
control system suffers from the following drawbacks.

\subsection{Scaling to Large Tunable Programs}
\label{subsec:scale-large-tunable}

The open-loop control system described in Section~\ref{sec:capri} works well for programs that are a few hundred lines
long and have five or six knobs. This holds true especially for the Bayesian network-based error model that was used as
a proxy function for $f_{e}$, as discussed in Section~\ref{sec:error-model}. There are many ways to model the error
probability distribution using a Bayesian network. The original Capri work used a simple model, where the output error E
depends directly on the settings of all of the knobs. Although this is simple, the size of the table for the output
error is exponential in the number of knobs. This was not a problem for the applications studied in the original Capri
paper. However, the control system may suffer from poor performance with applications that provide several ($\sim$ 100s) knobs
and therefore have a large space of knob settings.

There are possible ways to improve the performance of a Bayesian network-based cost model (or based on any machine
learning model), and to scale the open-loop controller. In the following, we discuss few opportunities:

\begin{itemize}
  
\item Reducing the size of the knob space: This can be achieved by (i) reducing the number of knobs that need to be
  controlled \emph{simultaneously}, a process that we call \emph{knob orthogonalization}, and (ii) reducing the number
  of settings for each knob.

  The first step is to exploit {\em phase behavior} in long-running programs~\cite{sherwood-asplos-2002,
    sherwood-micro-2003}. For example, a Barnes-Hut n-body code executes the following phases repeatedly: (i) build the
  spatial decomposition tree, (ii) compute the mass and center of gravity of each spatial partition, (iii) compute force
  on each particle, and (iv) update position and velocity of each particle. At any given point in the execution, the
  program is executing only one of these phases, so the overall control problem can be decomposed into a set of smaller
  control problems, one for each phase, thereby reducing the number of knobs that need to be controlled
  \emph{simultaneously}.

  The next step is to reduce the number of knobs by exploiting the 90/10 rule, which says that in most programs, more
  than 90\% of the execution time is spent in less than 10\% of the code. By ignoring knobs outside such ``hot''
  regions, it may be possible to obtain most of the benefits of optimal control without the effort of controlling every
  knob in a program. In Barnes-Hut for example, more than 90\% of the time is spent in the force computation phase, so
  it may be possible to ignore knobs in all other phases, at least for controlling computation time and energy.

  Once the number of knobs that need to be considered simultaneously has been minimized, reducing the number of knob
  \emph{settings} that need to be considered by the control system can be accomplished by using a mixture of
  coarse-grain and fine-grain knobs. If the program output changes relatively slowly with the value of a particular
  control variable, a coarse-grain knob with relatively few settings can be used to set the value of that variable,
  reducing the size of the search space for optimal knob settings. Profiling with test data can be used to determine the
  relative sensitivity of the output to particular control variables.

\item Reducing search time for optimal knob settings: Our current control algorithm sweeps the knob space to find
  optimal knob settings for a given input and desired quality guarantees. Although exhaustive search has worked well for
  the small-scale applications we have considered so far, it obviously does not scale to large numbers of knobs, so we
  will develop intelligent search algorithms to find optimal knob settings efficiently in a large knob space. As
  mentioned in Section~\ref{sec:capri}, we have experimented with the heuristic search strategy used in the
  Precimonious system~\cite{rubio-gonzalez-sc-2013}. The results showed, as one might expect, that Precimonious was
  significantly faster but found sub-optimal knob settings compared to our current exhaustive search strategy. In
  particular, for the OpenOrd application, Precimonious search got stuck in a local minimum that was sub-optimal. We
  will investigate search techniques that trade-off computing time for solution quality.

\item Scalable error models: The Bayesian error model in Capri has the virtue of being simple, but it does not scale
  since the size of the conditional probability distribution tables increases exponentially with the number of knobs.
  Abstractly, the error model is a function $f(v_1, v_2, ..., v_n, e_b)$ that maps a knob setting $(v_1, v_2, ..., v_n)$
  and an error bound $e_b$ to the corresponding probability defined in Problem formulation~\ref{opt:final}. The simple
  Bayesian model explicitly stores the probability for all combinations of knob settings and error bounds. In our
  studies, we have found that the probability function changes quite slowly as the knob settings are changed. Therefore,
  we might be able to usefully approximate the probability table by partitioning the knob space into subspaces and using
  a simple model like a linear model within each subspace. This is what a tool like M5 will do automatically if it is
  given the same training data as the Bayesian error model. We are investigating these model compression techniques.

\item Clustering inputs: Instead of building a single error model and cost model to handle all inputs, we can use
  clustering techniques~\cite{bishop-2006,mitchell-1997} to cluster the inputs into a set of classes where in each
  class, the error and cost behaviors are similar. For each class, we can build a separate quality and cost model using
  our approach. At runtime, a given input is first classified and then the corresponding models are used to set the
  knobs. This may improve both the accuracy and the scalability of both learning and querying the quality and cost
  models since the complexities of the models can be reduced by considering a subset of input scenarios. Clustering has
  been used successfully for auto-tuning in the Petabricks system~\cite{ansel-pldi-2009}. Automatic feature extraction and
  selection techniques may be useful for this problem; for example, they have been quite successful in the audio
  domain~\cite{mierswa-2005,monteiro-2010}.

\end{itemize}

\subsection{Closed-loop Control}
\label{subsec:closed-loop}

 Open-loop
control systems cannot adapt during execution to compensate for \emph{model error}. Such errors can be significant; for the SGD benchmark for example, the Capri control system does not find some low-cost points found by the oracle control because the Bayesian error model is overly conservative, as seen in  Figure~\ref{fig:modelAccuracy}.

The need to compensate for modeling errors, particularly in the context of complex systems, presents an opportunity for
\emph{closed-loop control}. In this approach, a function of the current system state and/or output is fed back as an
input to the control system so that system behavior can be optimized for subsequent computations. Closed-loop control
systems are generally applicable to a large class of iterative and streaming applications that have a notion of
``progress.'' For an iterative application, each iteration represents progress, and provides the control system with an
opportunity for correcting the difference between the current value of a system variable and the desired ``setpoint.''
For \emph{streaming} computations, the application processes a sequence of inputs and produces a sequence of outputs, so
the processing of successive inputs represents progress. Closed-loop control systems are well-studied in control theory,
and systematic techniques for designing controllers with provably desirable properties are well understood, especially
for linear, time-invariant systems. These techniques have proved to be adequate for simple cruise control systems in
cars, autopilots in aircraft, audio amplifiers, and basic process control systems in manufacturing.

Recently, there has been a surge in using closed-loop control to build adaptive software and hardware systems for
complex applications. However, there are several challenges: 1) building reasonable initial approximate cost and quality
models, 2) finding effective run-time metrics strongly correlated with cost and error/quality, 3) low overhead profiling
of these run-time metrics, and 4) updating knob settings, and cost and quality models efficiently. These ideas have been
explored in recent papers~\cite{baek-pldi-2010, meantime-atc-2016, jouleguard-sosp-2015, poet-2015, mithra-isca-2016,
  self-adaptive-software, slambench-2015, apex-2015} on adapting traditional control theory for use in computer applications. Some
of these systems consider a combination of system knobs, e.g. the number of cores used and their clock rate, in addition
to application knobs of the sort we use in Capri for open-loop control. However, existing systems typically use a
separate PID (proportional-integral-derivative) controller~\cite{astrom-2008} for each type of knob and employ ad-hoc
techniques to combine these into an overall system. PID controllers have the advantage of not needing system models but
because of this, they cannot ensure \emph{optimal} control; in addition, composing these controllers in ad hoc ways
limits the degree to which overall system behavioral properties can be guaranteed. They share these properties with
systems based on reinforcement learning~\cite{reinforcement-learning}. Such ad-hoc techniques are ill-suited for several emerging
class of application contexts such as exascale applications, which can execute for several days and which require tuning
of additional system-level knobs related to resource allocation such as load balancing and allocating cores~\cite{apex-2015}.

We propose to extend our model-based open-loop control framework to provide a systematic approach for designing
closed-loop controllers that integrate the use of system and application knobs to achieve predictable, desirable system
behavior. To that end, we will extend the strategy used in the established area of \emph{Model-Predictive Control} (MPC)
for traditional control systems~\cite{seduction-2006, camacho-1997}. Traditional MPC systems are used to design relatively complex process control systems for industrial plants, and can be more effective than simple PID controllers. Unlike PID controllers, they are based on specific, closed-form dynamics models of the processes being controlled. Based on these models, an explicit closed form objective function describing the desired system behavior as well as explicit closed-form constraints on the range of behaviors allowed can be expressed. This results in a formulation of the control problem as a constrained optimization problem, where the behavioral objective function is optimized subject to the specified constraints. Note that this is similar to our formulation for the open-loop problem. In traditional MPC systems, these functions are closed-form continuous functions to be optimized over an infinite time period. To make the problem computationally tractable, a finite time horizon is imposed and the optimal trajectory of the control settings over that time horizon is computed. Since in the real control system, knobs are set at discrete time intervals, the setting computed by the optimizer for the first time interval is used by the controller. The optimization step is then repeated for the given horizon, and the first step of the resulting trajectory actually used, and so on. Of course, this comes at the cost of more expensive computation per time step than for PID controllers. In principle, the traditional MPC method can be extended to non-linear systems, although the resulting nonlinear optimization problems may be too expensive to solve, for real-time use, using traditional methods.

In the complex systems we wish to consider, we know of no closed-form models for the cost and quality functions, but we
can model these using machine learning techniques as described in Section~\ref{sec:capri}. These will constitute the
initial approximate cost and quality models for the proposed closed-loop control system. We are currently working on
finding effective run-time metrics strongly correlated with cost and error/quality for the applications discussed in
Section~\ref{sec:capri}. We are analyzing \underline{S}imultaneous \underline{L}ocalization \underline{A}nd
\underline{M}apping (SLAM) applications to determine sensitivity to platform knobs as well as the best application knobs
to use in trading off accuracy for computation time and energy savings~\cite{slambench-2015}. We are also exploring the
incorporation of our MPC-based controllers into systems like APEX~\cite{apex-2015} for controlling exascale computations
(the current APEX system uses a simple proportional controller to control the number of cores assigned to a computation,
for example). We believe these kinds of real-world applications are a rich source of interesting problems for our
proposed extensions.

We are also working on methods for incrementally updating optimal knob settings using the models and feedback
information about the state of the computation. If this is done at each iteration of a streaming computation, we can see
that this fits the model of MPC control with a time horizon for optimization of a single time-step. The final step will
be to incorporate model updates into the system, as is done in approaches like Kalman filters in traditional control
theory~\cite{astrom-2008}. In our current open-loop control system, we do not take into account the results of previous
computations to refine the models constructed during the initial training. For online control, it may be desirable to
incorporate some kind of model refinement so that subsequent optimization steps improve in quality. A related goal is to
develop techniques for guaranteeing that our systems converge to desired behaviors using this approach. Finally, we will
develop techniques for multi-time-step optimization to provide better results on appropriate problems such as
recognizing and tracking the motion of objects with multi-model sensors.


\section{Making Capri Extensible}
\label{sec:impl}

Section~\ref{sec:extensions} discusses possible extensions to Capri~\cite{capri-asplos-2016}. We are actively working on
tailoring the existing \capri implementation to integrate extensions. This section describes our new \capri implementation in
detail, and uses applications from three varied domains to demonstrate the effectiveness of and regression test the new
\capri implementation.

\subsection{Applications}
\label{sec:applications}

We evaluate the new \capri system on three complex applications: (i) \gem, the graph partitioner for social
networks~\cite{whang-icdm-2012} which was introduced earlier, (ii) a radar processing
application~\cite{hoffmann-tpds-2012} written by researchers at the University of Chicago, and (iii) SLAMBench, an open
source tool designed to assist in the development of simultaneous localisation and mapping (SLAM)
algorithms~\cite{slambench-2015}. The code for \gem was modified by us to permit control of approximation, while Radar
and SLAMBench were already setup for control.

\paragraph*{Error/Quality definition.} To compute the error/quality of the output, we require the user to provide a
distance function that quantifies the difference between an approximate execution and a reference execution for a given
input. The reference execution can be the exact execution if such a thing exists or the best execution in the knob space
for that input. The error is defined as a normalized version of this distance
\[\text{Error} = (d - d_{min})/(d_{max} - d_{min}) \] where $d$, $d_{max}$ and $d_{min}$ represent the distance for a
execution, the maximum distance and the minimum distance over the knob space for the same input. The distance function
is application-specific.

\subsubsection{GEM}

GEM~\cite{whang-icdm-2012} is a graph clustering algorithm for social networks.

\textbf{Knobs:} There are two components; both use a weighted kernel k-means algorithm and have a knob controlling the
number of iterations. Each knob can be set to one of $40$ levels. All input graphs are partitioned into $100$ clusters
in our experiments.

\textbf{Error metric:} The output of GEM is the cluster assignment of each node in the graph. There is a standard way to
measure the quality of graph clustering, using the notion of a \emph{normalized cut}, which is defined as follows:
\[
 \scriptsize \sum_{k=1}^{N} \sum_{i=1, i \neq k}^{N} \text{edges}(C_{k}, C_{i})/\text{edges}(C_{k})
\]
where $N$ is the number of clusters, $\text{edges}(C_{k}, C_{i})$ denotes the number of edges between cluster $k$ and cluster $i$, and $\text{edges}(C_{k})$ denotes the edges inside cluster $k$

The distance function computes the difference of the normalized cut given two clustering assignments. The reference execution is the execution achieving the smallest normalized cut.

\textbf{Input features for modeling cost:} the number of vertices in the graph, the number of edges and the number of clusters.

\subsubsection{Radar}

We used a radar processing application~\cite{hoffmann-tpds-2012} developed by Hank Hoffmann at the University of
Chicago. Unlike the other applications, this code was already instrumented with knobs, so we used it out
of the box as a blind test for our system. This code is a pipeline with four stages. The first stage (LPF) is a low-pass
filter to eliminate high-frequency noise. The second stage (BF) does beam-forming which allows a phased array radar to
concentrate in a particular direction. The third stage (PC) performs pulse compression, which concentrates energy. The
final stage is a constant false alarm rate detection (CFAR), which identifies targets.

\textbf{Knobs:} The application supports four knobs. The first two knobs change the decimation ratios in the finite
impulse response filters that make up the LPF stage. The third knob changes the number of beams used in the beam former.
The fourth knob changes the range resolution. The application can have 512 separate configurations using these four
knobs.

\textbf{Error metrics:} The signal-to-noise ratio (SNR) is used to measure the quality of the detection. The reference
execution is the one achieving the highest SNR.

\textbf{Input features for modeling cost:} No input features are used in this application.

\subsubsection{SLAMBench}

SLAMBench is an open source tool designed to assist in the development of simultaneous localisation and mapping (SLAM)
algorithms, and evaluation of platforms for implementing those algorithms~\cite{slambench-2015}. It runs on the Linux
operating system, and has been used on X86 and ARM along with various GPUs, from high-end to mobile devices. SLAMBench combines
a framework for quantifying quality-of-result with instrumentation of execution time and energy consumption. It contains
a KinectFusion~\cite{kinectfusion} implementation in C++, OpenMP, OpenCL and CUDA. It offers a platform for a broad
spectrum of future research in jointly exploring the design space of algorithmic and implementation-level optimizations.

\textbf{Knobs:} The application supports several algorithmic-level knobs~\cite{bodin-2016}, such as, volume resolution,
iterative closest point (ICP) threshold, etc. To minimize the search space over the set of all possible knob
combinations, we vary only those knobs that seem to have a high correlation with the run-time performance and tracking.
We used the following four algorithmic parameters as knobs:
\begin{itemize}
\item Compute size ratio - The fractional depth image resolution used as input. As an example, a value of 8 means that
  the raw frame is resized to one-eighth resolution.
\item ICP threshold - The threshold for the iterative closest point (ICP) algorithm used during the tracking phase in KinectFusion.
\item $\mu$ distance - The output volume of KFusion is defined as a truncated signed distance function
  (TSDF)~\cite{kinectfusion}. Every volume element (voxel) of the volume contains the best likelihood distance to the
  nearest visible surface, up to a truncation distance denoted by the parameter $\mu$.
\item Volume resolution - The resolution of the scene being reconstructed. As an example, a 64x64x64 voxel grid captures
  less detail than a 256x256x256 voxel grid.
\end{itemize}

\textbf{Error metrics:} The KinectFusion algorithm reports the absolute trajectory error (ATE) in meters after
processing an input. The ATE measures accuracy, and represents the precision of the computation. Acceptable values are
in the range of few centimeters.

\textbf{Input features for modeling cost:} An input in SLAMBench is a trajectory, which is sequence of depth images. We
have defined the following features that can be extracted from a given trajectory: mean and standard deviation of the
depth values in a frame, and mean and standard deviation of differences in depth values between successive frames. The
first two features track the variation among pixels in a single frame, while the second pair of features aim to capture
the variation in depth values across two images. In other words, it tries to capture the ``burstiness'' between two
successive frames.

\subsection{Implementation}

\paragraph*{Environment.} \Capri has been implemented and tested with Python v3.5.

\paragraph*{M5 model.}
The \code{Cubist}\footnote{\url{http://rulequest.com/cubist-info.html}} application implements the M5~\cite{quinlan-92}
machine learning model. Training a Cubist/M5 model requires a schema file which lists the independent and the dependent
variables, and a file containing data points that is used for training. After training, Cubist/M5 generates a set of
piecewise-linear rule-based models that balance the need for accurate prediction against the requirements of
intelligibility. Cubist/M5 models generally give better results than those produced by simple techniques such as
multivariate linear regression, while also being easier to understand than the more complex neural networks. Cubist/M5
scales well to hundreds of attributes.

\Capri uses the \code{Cubist}\footnote{\url{https://cran.r-project.org/web/packages/Cubist/index.html}} package
available for the \code{R} programming language. We have written a Python package wrapper for interfacing with the
Cubist \code{R} package. Our new \capri implementation is modular, which makes it easy to replace the M5 model with other
machine learning models.

\paragraph*{Source structure.}

The \capri source is divided into the follow directories:
\begin{itemize}
\item \code{lib} - Contains the source for Cubist \code{R} module, and a Python wrapper for interfacing
  with \capri.
\item \code{scripts} - Contains scripts for helping with running applications and parsing the output results.
\item \code{src} - This directory contains Python modules that implement the control algorithm in \capri.
\end{itemize}

\paragraph*{Running \Capri.}

\Capri can be run with Python version $\geq$ 3.5, and requires the following Python packages: \code{numpy},
\code{psutil}, \code{overrides}, \code{matplotlib}, and \code{ordered\_set}. These packages can be installed by invoking
the following command if required: \lstset{style=bash}
\begin{lstlisting}{style=mybash}
  pip3 install --upgrade numpy psutil overrides matplotlib ordered_set
\end{lstlisting}

Since each application has a unique set of knobs and a different range of values, therefore an user of the \capri system
needs to list the details about the knobs and their range of values in a configuration file. The \capri source provides
configuration files for several applications that we have used. The following snippet shows an example for the \gem
application:
\begin{verbatim}
[FIXED]

PBS = (1.0;0.05;-0.05)
EBS = (0.05;1;+0.05)
TRAIN_RATIO = 0.75
ACCURATE_KNOBS = {'iter1': '40', 'iter2': '40'}

[KNOBS]

NUM_FIRST_ITER = (1;40;+1) # iter1
NUM_SECOND_ITER = (1;40;+1) # iter2
\end{verbatim}

\noindent
The configuration section \code{FIXED} lists experimental settings that are common to all applications. \code{PBS} and
\code{EBS} stand for the acceptable probability and error bounds, as discussed in the final problem
formulation~\ref{opt:final} in Section~\ref{sec:problem}. \code{TRAIN\_RATIO} specifies the proportion of the
experimental data to be used in training the M5 models, the rest of the input data is used for prediction.
\code{ACCURATE\_KNOBS} specifies the knob configurations that compute the most accurate output, which is used to compute
the ``golden value'' (i.e., the most accurate output) and scale the error.

Given a configuration file for an application, we have automated all the steps involved in running the \capri control
system with the application. Executing the control system involves four steps:
\begin{itemize}
\item \code{run} - Run the application with different knob settings to generate experimental data to be then used for offline machine learning and for prediction. This task does not
  depend on other tasks, and can be run independently. Note that running this task over all possible knob settings can
  take a long time (i.e., several hours to several days depending on the application).
\item \code{stats} - Process a set of experimental results to collect statistics. This task depends on output generated
  by a prior \code{run} task.
\item \code{predict} - Train the M5 models and compute the feasible region for a given constraint of error bound
  ($\epsilon$) and probability bound ($\pi$) (Section~\ref{sec:problem}). This task depends on the \code{run} task.
\item \code{result} - Find the optimal knob setting that minimizes the objective function and meets the constraints set
  in Problem formulation~\ref{opt:final}. It also generates plots and speedups to help compare the performance of the
  \capri control system. This task depends on the \code{stats} and the \code{predict} tasks.
\end{itemize}

In the following, we show a sample invocation of the \capri control system.

\begin{lstlisting}{style=bash}
capri --bench=gem --input=all --outputDir=gem-full --tasks=run,stats,predict,result
\end{lstlisting}

To know more about different options to \capri, use
\begin{lstlisting}{style=bash}
  capri -h
\end{lstlisting}

\paragraph*{Extending \capri with new applications.}

It is straightforward to add support for new applications to \capri. A user of the \capri system needs to provide the
following information:
\begin{itemize}
\item Implement an application-specific module under \code{apps} in the \code{src} directory. The \capri user should
  implement how to compute the cost and the error for the application. Please refer to existing applications for
  reference.
\item Provide a configuration file for the application.
\item Use \capri to run the application with different knob settings, and then use the controller to predict optimal
  knobs for any given performance metric and quality constraints.
\end{itemize}


\section{Evaluation}
\label{sec:eval}

In this section, we describe results using our new implementation of the \capri control system with \gem, Radar, and
SLAMBench. For each benchmark, we collected a set of inputs as shown in Table~\ref{table:inputs}. We would have liked to
have more training inputs for \gem, and we are currently investigating other sources of getting new inputs for
SLAMBench. To evaluate the error and cost models, inputs were randomly partitioned into a training and a test suite
based on the \code{TRAIN\_RATIO}.

\begin{table}
\footnotesize
\centering
  \begin{tabular}{@{}lrrrl@{}}
    Benchmark & \#Total & \#Train (75\%) & \#Test (25\%) & Source\\ \midrule
    GEM & 43 & 33 & 10  &  \cite{stanford-snap, zafarani-liu-2009}\\
    Radar & 128 & 96 & 32 & synthetic \\
    SLAMBench & 12 & 9 & 3 & ~\cite{slambench-2015}
  \end{tabular}
\caption{Inputs for benchmarks. Inputs are randomly divided into training set and testing set.}
\label{table:inputs}
\end{table}


\begin{figure}[t]
  \centering
  \subfloat[\Gem]{\includegraphics[height=4cm, width=0.33\linewidth]{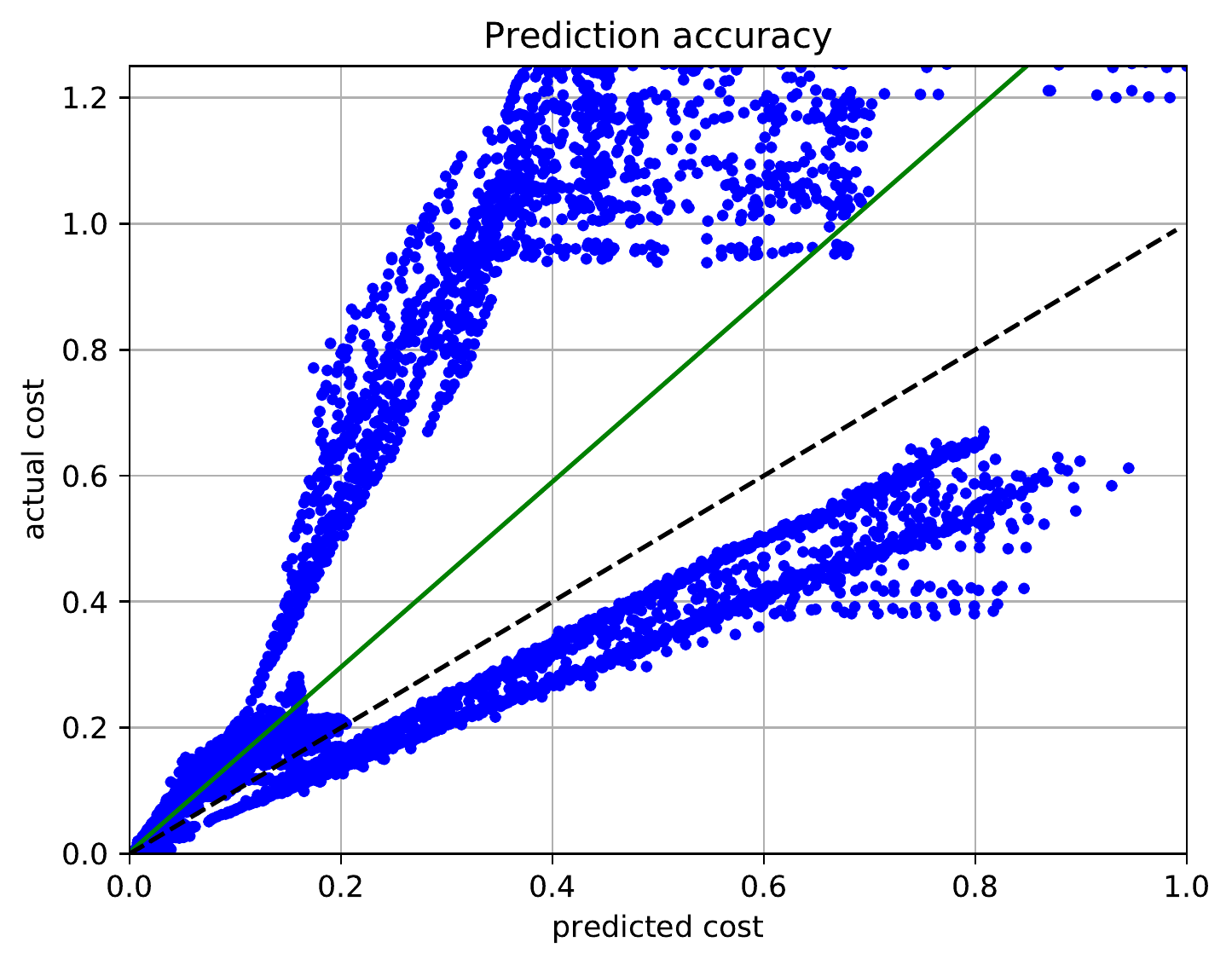}}
  \subfloat[Radar application]{\includegraphics[height=4cm, width=0.33\linewidth]{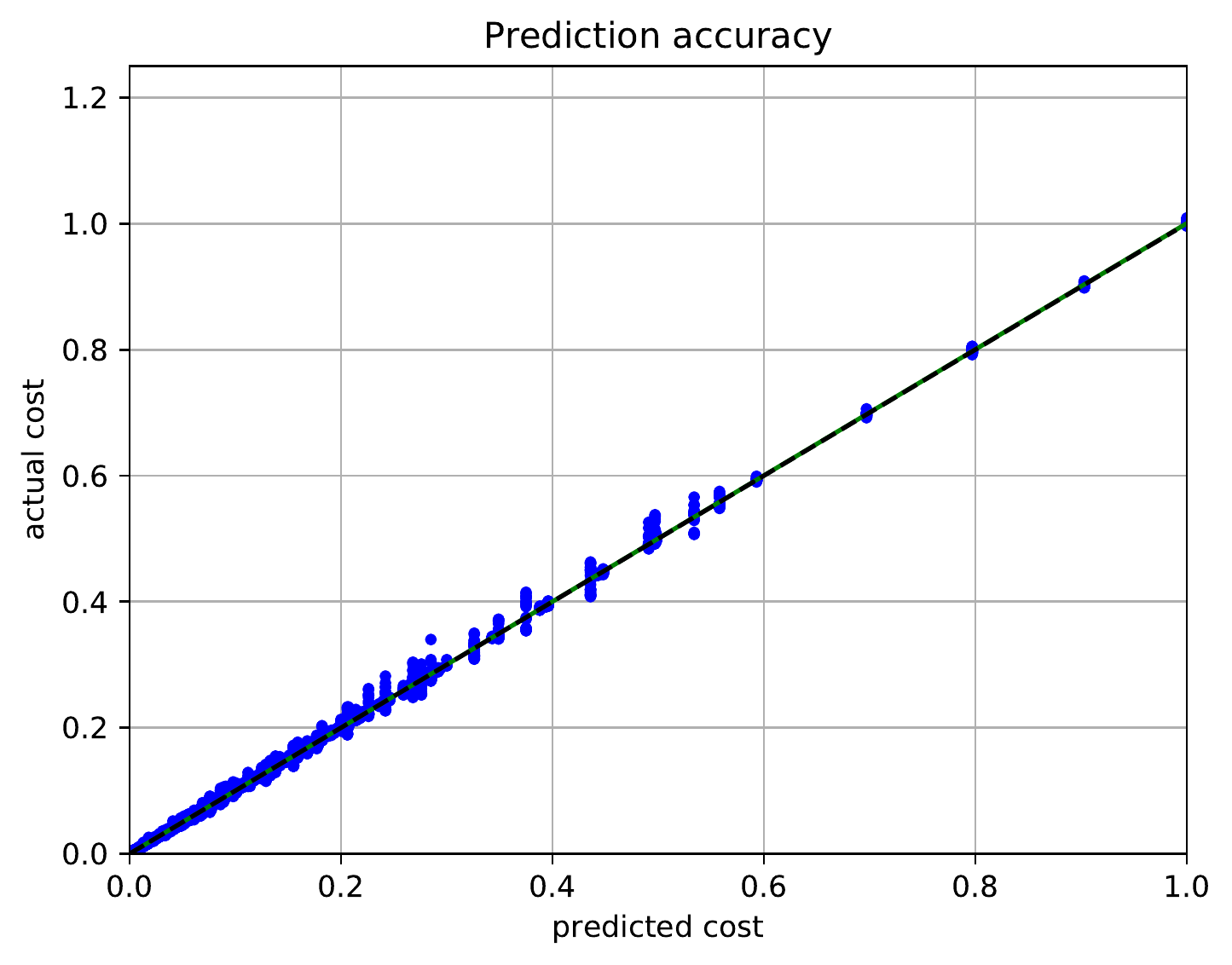}}
  \subfloat[SLAMBench]{\includegraphics[height=4cm, width=0.33\linewidth]{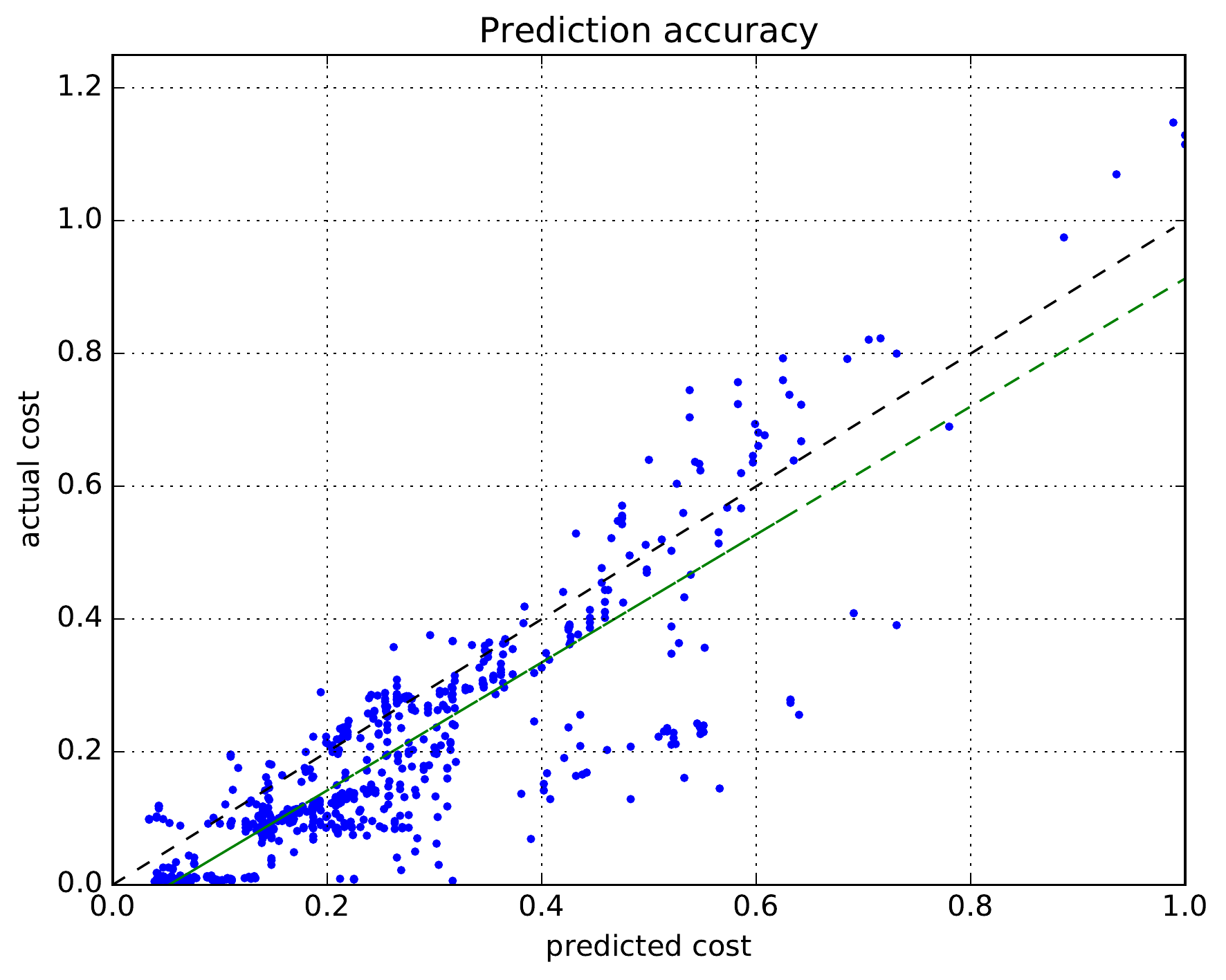}}
  \caption{Accuracy of the cost model with the new implementation of \capri.}
  \label{fig:cost-accuracy}
\end{figure}

\subsection{Evaluation of the Cost and Error Models}
\label{sec:eval-cost-error}

We regression test our new implementation of \capri by comparing the performance of the M5 cost and error models with
prior results (Section~\ref{sec:capri:results}). Figure~\ref{fig:cost-accuracy} shows the accuracy of the M5 cost model.
As in Figure~\ref{fig:modelAccuracy}, the black line represents the \code{y=x} line which captures perfect prediction
behavior. The green line shows linear regression for the given data points. From Figures~\ref{fig:modelAccuracy}
and~\ref{fig:cost-accuracy}, it is obvious that the behavior of the new M5 cost model closely matches the earlier
result.

\begin{figure}[t]
  \centering
  \subfloat[\Gem]{\includegraphics[height=4cm, width=0.33\linewidth]{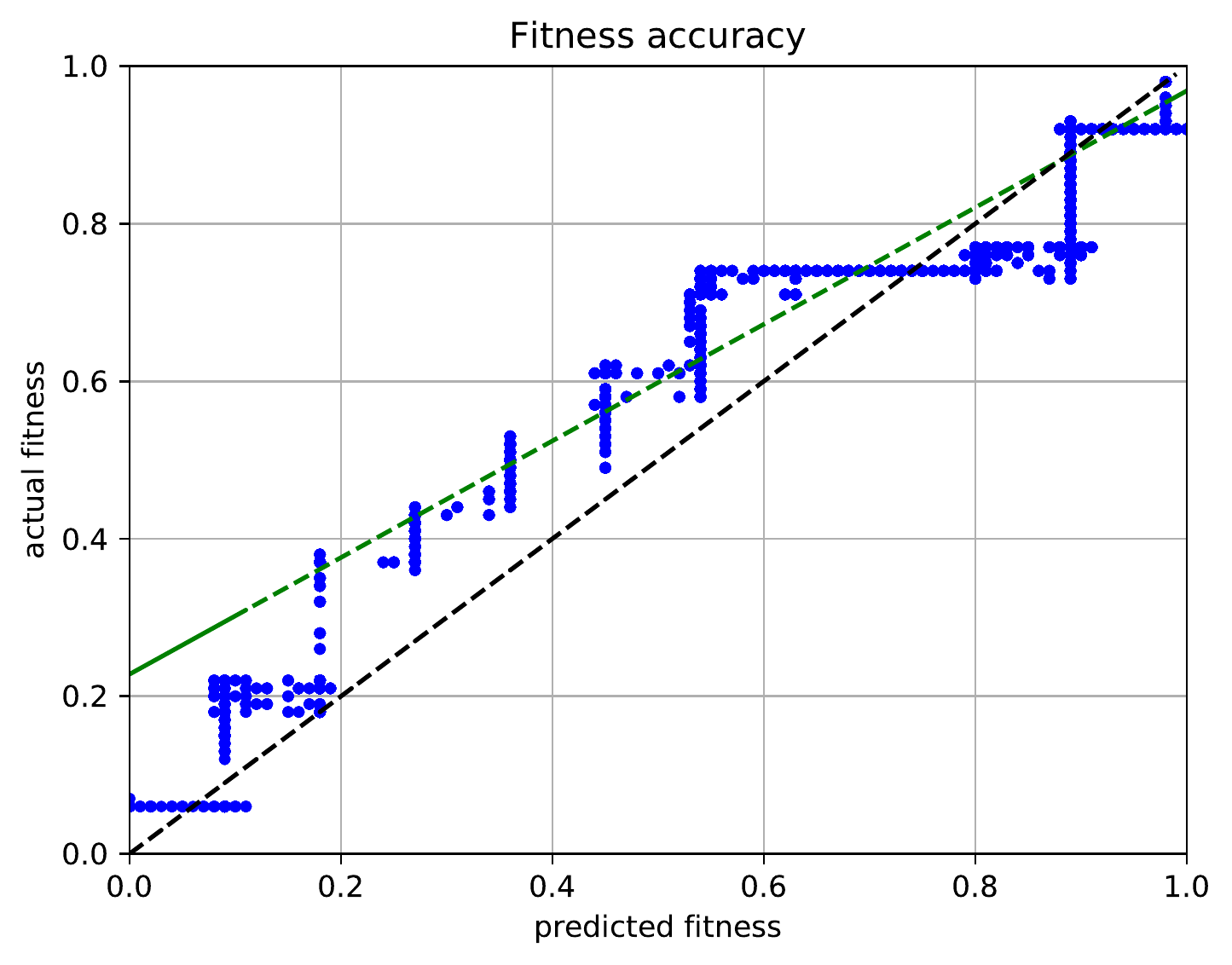}}
  \subfloat[Radar application]{\includegraphics[height=4cm, width=0.33\linewidth]{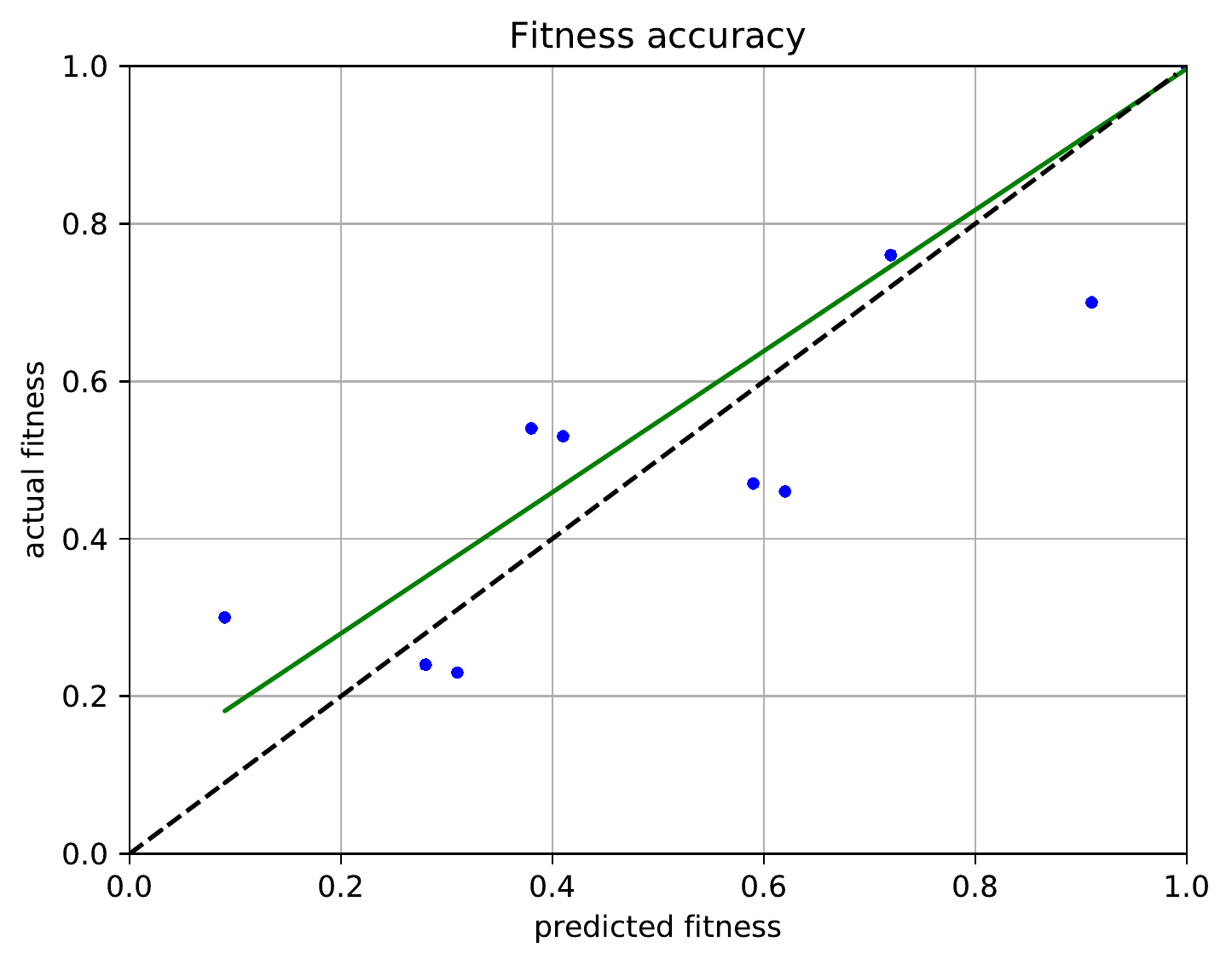}}
  \subfloat[SLAMBench]{\includegraphics[height=4cm, width=0.33\linewidth]{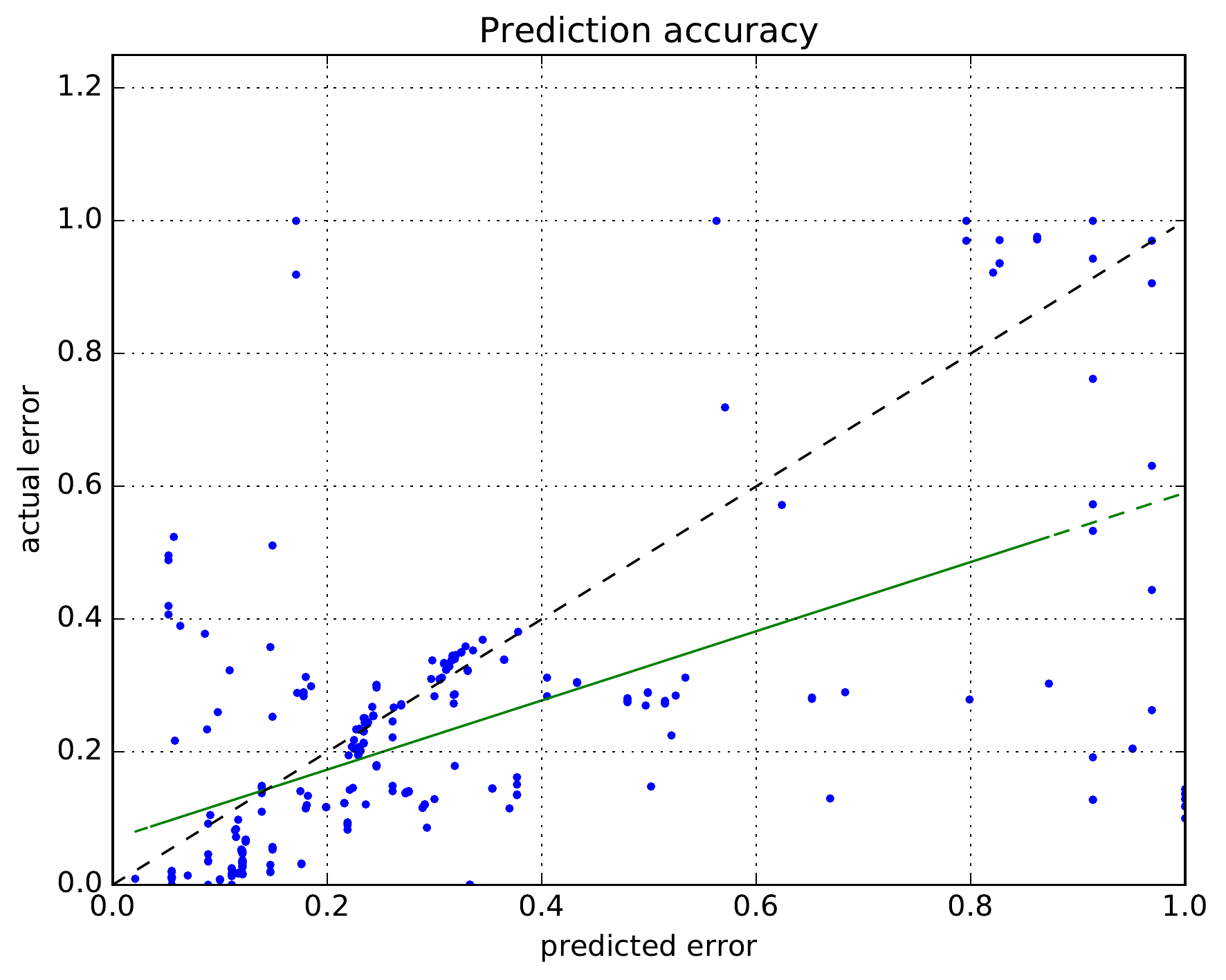}}
  \caption{Accuracy of the error model with the new implementation of \capri.}
  \label{fig:fitness-accuracy}
\end{figure}

Prior published work on \capri used Bayesian network for modeling error~\cite{capri-asplos-2016}. Unlike prior work, our
reimplementation uses M5 for modeling error. Figure~\ref{fig:fitness-accuracy} shows the accuracy of the M5 cost model.
Evaluating the accuracy of the error model is more involved than the cost model, since the error bound needs to be met
probabilistically over an ensemble of inputs. We simulate that by tracking the proportion of the inputs for which the
\capri control system's predictions meet the given error bound. The black and green lines in the figure have the same
meaning as in Figure~\ref{fig:cost-accuracy}. From Figures~\ref{fig:modelAccuracy} and~\ref{fig:fitness-accuracy}, we
see that predictions with an M5 model are within a reasonable match of predictions with a Bayesian network. Fitness
predictions for \bench{SLAMBench} are wayward, we believe this is due to lack of sufficient training data and a poor
choice of the error function (based on absolute trajectory error). We are investigating ways to fix this problem with
SLAMBench. In particular, we are looking into how to use the RGB-D SLAM dataset from
TUM\footnote{\url{http://vision.in.tum.de/data/datasets/rgbd-dataset}} and to generate new trajectories.

\begin{table}[t]
  \centering
  \begin{tabular}{l|rrrrrr|rrrrrr|rrrrrr}
    \toprule
    & \multicolumn{6}{c}{\Gem} & \multicolumn{6}{|c|}{Radar} & \multicolumn{6}{c}{SLAMBench} \\
    \hline
    $\pi\parallel\epsilon$ & 0.0 & 0.1 & 0.2 & 0.3 & 0.4 & 0.5 & 0.0 & 0.1 & 0.2 & 0.3 & 0.4 & 0.5 & 0.0 & 0.1 & 0.2 & 0.3 & 0.4 & 0.5 \\
    \midrule
    1.0 & NA & NA  & 1.3 & 1.7 & 2.2 & 2.3 & 1 & 1 & 1.1 & 1.3 & 1.4 & 1.4 & NA & NA  & NA  & 1.6 & 1.6 & 1.5 \\
    0.9 & NA & 1.1 & 2.1 & 2.8 & 3.2 & 3.7 & 1 & 1 & 1.1 & 1.3 & 1.4 & 1.4 & NA & NA  & 1.2 & 1.6 & 1.6 & 1.5 \\
    0.8 & NA & 1.6 & 2.4 & 3.2 & 3.6 & 4.0 & 1 & 1 & 1.1 & 1.3 & 1.4 & 1.4 & NA & 1.2 & 1.5 & 1.5 & 1.5 & 1.5 \\
    0.7 & NA & 1.8 & 2.8 & 3.5 & 3.9 & 4.4 & 1 & 1 & 1.1 & 1.3 & 1.4 & 1.4 & NA & 1.2 & 1.5 & 1.5 & 1.5 & 1.5 \\
    0.6 & NA & 2.3 & 3.2 & 3.6 & 4.0 & 4.8 & 1 & 1 & 1.1 & 1.3 & 1.4 & 1.4 & NA & 1.3 & 1.5 & 1.5 & 1.5 & 1.5 \\
    0.5 & NA & 2.6 & 3.4 & 3.9 & 4.1 & 4.9 & 1 & 1 & 1.1 & 1.3 & 1.4 & 1.4 & NA & 1.3 & 1.5 & 1.5 & 1.5 & 1.5 \\
    \bottomrule
  \end{tabular}
  \caption{Speedups of the tuned programs for a subset of constraint space. }
  \label{table:cost-speedups}
\end{table}

\subsection{Speedups}
\label{sec:speedups}

The original \capri work shows that the time for control is relatively small compared to the time taken by the
applications to run. In our reevaluation, we have not measured the proportion of the time taken by the control algorithm
to run compared to the applications. But we evaluate \emph{speedup} to sanity check the performance of the new \capri
implementation. Speedup is defined as ratio of the running time at a particular knob setting to the running time with
the knobs set for maximum quality.

Table~\ref{table:cost-speedups} shows speedups for each application for $\epsilon$ values between 0 and 0.5 and $\pi$
values between 0.5 and 1.0 (we show only a portion of the overall constraint space for simplicity). Each entry gives the
average speedup over all test inputs for the knob settings found by the control algorithm based on exhaustive search,
given $(\epsilon, \pi)$ constraints in the intervals specified by the row and column indices.

Speedups depend on the application and the $(\epsilon \pi)$ constraints. For each application, the top-left corner of
the constraint space is the ``hard'' region since the error must be low with high probability. The knob settings must be
at or close to maximum, and speedup will be limited. Table entries marked ``NA'' show where the control system was
unable to find any feasible solution for these hard constraints. In contrast, the bottom-right corner of the constraint
space is the ``easier'' region, so one would expect higher speedups. This is seen with all the applications. Overall, we see
that controlling the knobs in these applications can yield significant speedups in running time.

\subsection{Inversions}
\label{subsec:inversions}

\begin{table}[t]
  \centering
  \begin{tabular}{l|rrrrrr|rrrrrr|rrrrrr}
    \toprule
    & \multicolumn{6}{c}{\Gem} & \multicolumn{6}{c}{Radar} & \multicolumn{6}{c}{SLAMBench} \\
    \hline
    $\pi\parallel\epsilon$ & 0.0 & 0.1 & 0.2 & 0.3 & 0.4 & 0.5 & 0.0 & 0.1 & 0.2 & 0.3 & 0.4 & 0.5 & 0.0 & 0.1 & 0.2 & 0.3 & 0.4 & 0.5 \\
    \midrule
    1.0 & NA & NA & F & T & T & T & F & F & F & F & F & F & NA & NA & NA & F & F & T \\
    0.9 & NA & F  & F & F & T & F & F & F & F & F & F & F & NA & NA & F  & F & F & T \\
    0.8 & NA & F  & F & T & T & F & F & F & F & F & F & F & NA & T  & F  & T & T & T \\
    0.7 & NA & F  & T & F & T & F & F & F & F & F & F & F & NA & T  & F  & T & T & T \\
    0.6 & NA & F  & T & F & F & T & F & F & F & F & F & F & NA & T  & F  & T & T & T \\
    0.5 & NA & F  & T & T & F & F & F & F & F & F & F & F & NA & F  & T  & T & T & T \\
    \bottomrule
  \end{tabular}
  \caption{Inversion of the tuned programs for a subset of constraint space. }
  \label{table:inversions}
\end{table}

The cost and error models in \capri are used only to rank knob settings in the feasible region, so more accurate models
do not necessarily give better solutions to the control problem even if the predictions of the machine learning models are
close to accurate. We say an \emph{inversion} has occurred for a given constraint of $\epsilon$ and $\pi$ when the
knob setting predicted by \capri does not match with the knob settings identified with an oracle. We evaluated the
number of inversions that happened with the M5 model in \capri, by comparing whether the predicted knob settings matched
with the knob settings predicted using an oracle for a given $\epsilon$ and $\pi$ constraint.
Table~\ref{table:inversions} show the proportion of inversions that occurred with the different applications. We denote
an inversion has occurred with \code{T}, otherwise the entry contains \code{F}. The table shows that the machine
learning models in \capri perform reasonably well.


\section{Conclusion}
\label{sec:conclusion}

Although there is a large body of work on using approximate computing to reduce computation time as well as power and
energy requirements, little is known about how to control approximate programs in a principled way. Previous work on
approximate computing has focused either on showing the feasibility of approximation or on controlling streaming
programs in which error estimates for one input can be used to reactively control error for subsequent inputs.

In this paper, we addressed the problem of controlling tunable approximate programs, which have one or more knobs that
can be changed to vary the fidelity of the output of the approximate computation. We showed how the proactive control
problem for tunable programs can be formulated as an optimization problem, and then gave an algorithm for solving this
control problem by using error and cost models generated using machine learning techniques. Our experimental results
show that this approach performs well on controlling tunable approximate programs.

We extend prior published work called \capri to make the new control system scale to hundreds of knobs, and to provide optimal
control for streaming programs. For controlling streaming programs, we propose to solve a closed-loop control system
with model-predictive control. We showed initial results with our new implementation of \capri to regression test the
system.


{
  \bibliographystyle{abbrv}
  \bibliography{paper}

\begin{thebibliography}{10}

\bibitem{ansel-pldi-2009}
J.~Ansel, C.~Chan, Y.~L. Wong, M.~Olszewski, Q.~Zhao, A.~Edelman, and
  S.~Amarasinghe.
\newblock {PetaBricks: A Language and Compiler for Algorithmic Choice}.
\newblock In {\em Proceedings of the 30\textsuperscript{th} ACM SIGPLAN
  Conference on Programming Language Design and Implementation}, PLDI '09,
  pages 38--49, New York, NY, USA, 2009. ACM.

\bibitem{ansel-cgo-2011}
J.~Ansel, Y.~L. Wong, C.~Chan, M.~Olszewski, A.~Edelman, and S.~Amarasinghe.
\newblock {Language and Compiler Support for Auto-Tuning Variable-Accuracy
  Algorithms}.
\newblock In {\em Proceedings of the 9\textsuperscript{th} Annual IEEE/ACM
  International Symposium on Code Generation and Optimization}, CGO '11, pages
  85--96, Washington, DC, USA, 2011. IEEE Computer Society.

\bibitem{astrom-2008}
K.~J. {\AA}str{\"o}m and R.~M. Murray.
\newblock {\em {Feedback Systems: An Introduction for Scientists and
  Engineers}}.
\newblock Princeton University Press, Princeton, NJ, USA, 2008.

\bibitem{baek-pldi-2010}
W.~Baek and T.~M. Chilimbi.
\newblock {Green: A Framework for Supporting Energy-Conscious Programming using
  Controlled Approximation}.
\newblock In {\em Proceedings of the 31\textsuperscript{st} ACM SIGPLAN
  Conference on Programming Language Design and Implementation}, PLDI '10,
  pages 198--209, New York, NY, USA, 2010. ACM.

\bibitem{bienia-2011}
C.~Bienia.
\newblock {\em {Benchmarking Modern Multiprocessors}}.
\newblock PhD thesis, Princeton, NJ, USA, Jan. 2011.
\newblock AAI3445564.

\bibitem{bishop-2006}
C.~M. Bishop.
\newblock {\em {Pattern Recognition and Machine Learning (Information Science
  and Statistics)}}.
\newblock Springer-Verlag New York, Inc., Secaucus, NJ, USA, 2006.

\bibitem{bodin-2016}
B.~Bodin, L.~Nardi, M.~Z. Zia, H.~Wagstaff, G.~Sreekar~Shenoy, M.~Emani,
  J.~Mawer, C.~Kotselidis, A.~Nisbet, M.~Lujan, B.~Franke, P.~H. Kelly, and
  M.~O'Boyle.
\newblock {Integrating Algorithmic Parameters into Benchmarking and Design
  Space Exploration in 3D Scene Understanding}.
\newblock In {\em Proceedings of the 2016 International Conference on Parallel
  Architectures and Compilation}, PACT '16, pages 57--69, New York, NY, USA,
  2016. ACM.

\bibitem{bottou-2010}
L.~Bottou.
\newblock {Large-Scale Machine Learning with Stochastic Gradient Descent}.
\newblock In Y.~Lechevallier and G.~Saporta, editors, {\em Proceedings of the
  19\textsuperscript{th} International Conference on Computational Statistics
  (COMPSTAT'2010)}, pages 177--186, Heidelberg, 2010. Physica-Verlag HD.

\bibitem{camacho-1997}
E.~F. Camacho and C.~A. Bordons.
\newblock {\em {Model Predictive Control in the Process Industry}}.
\newblock Springer-Verlag New York, Inc., Secaucus, NJ, USA, 1997.

\bibitem{campanoni-cgo-2015}
S.~Campanoni, G.~Holloway, G.-Y. Wei, and D.~Brooks.
\newblock {HELIX-UP: Relaxing Program Semantics to Unleash Parallelization}.
\newblock In {\em Proceedings of the 13\textsuperscript{th} Annual IEEE/ACM
  International Symposium on Code Generation and Optimization}, CGO '15, pages
  235--245, Washington, DC, USA, 2015. IEEE Computer Society.

\bibitem{carbin-oopsla-2013}
M.~Carbin, S.~Misailovic, and M.~C. Rinard.
\newblock {Verifying Quantitative Reliability for Programs That Execute on
  Unreliable Hardware}.
\newblock In {\em Proceedings of the 2013 ACM SIGPLAN International Conference
  on Object Oriented Programming Systems Languages \& Applications}, OOPSLA
  '13, pages 33--52, New York, NY, USA, 2013. ACM.

\bibitem{chaudhuri-2012}
S.~Chaudhuri, S.~Gulwani, and R.~Lublinerman.
\newblock {Continuity and Robustness of Programs}.
\newblock {\em Communications of the ACM}, 55(8):107--115, Aug. 2012.

\bibitem{smooth-interpretation}
S.~Chaudhuri and A.~Solar-Lezama.
\newblock {Smooth Interpretation}.
\newblock In {\em Proceedings of the 31\textsuperscript{st} ACM SIGPLAN
  Conference on Programming Language Design and Implementation}, PLDI '10,
  pages 279--291, New York, NY, USA, 2010. ACM.

\bibitem{ding-pldi-2015}
Y.~Ding, J.~Ansel, K.~Veeramachaneni, X.~Shen, U.-M. O'Reilly, and
  S.~Amarasinghe.
\newblock {Autotuning Algorithmic Choice for Input Sensitivity}.
\newblock In {\em Proceedings of the 36\textsuperscript{th} ACM SIGPLAN
  Conference on Programming Language Design and Implementation}, PLDI '15,
  pages 379--390, New York, NY, USA, 2015. ACM.

\bibitem{esmaeilzadeh-asplos-2012}
H.~Esmaeilzadeh, A.~Sampson, L.~Ceze, and D.~Burger.
\newblock {Architecture Support for Disciplined Approximate Programming}.
\newblock In {\em Proceedings of the Seventeenth International Conference on
  Architectural Support for Programming Languages and Operating Systems},
  ASPLOS XVII, pages 301--312, New York, NY, USA, 2012. ACM.

\bibitem{esmaeilzadeh-micro-2012}
H.~Esmaeilzadeh, A.~Sampson, L.~Ceze, and D.~Burger.
\newblock {Neural Acceleration for General-Purpose Approximate Programs}.
\newblock In {\em Proceedings of the 2012 45\textsuperscript{th} Annual
  IEEE/ACM International Symposium on Microarchitecture}, MICRO-45, pages
  449--460, Washington, DC, USA, 2012. IEEE Computer Society.

\bibitem{fang-taco-2014}
S.~Fang, Z.~Du, Y.~Fang, Y.~Huang, Y.~Chen, L.~Eeckhout, O.~Temam, H.~Li,
  Y.~Chen, and C.~Wu.
\newblock {Performance Portability Across Heterogeneous SoCs Using a
  Generalized Library-Based Approach}.
\newblock {\em ACM Transactions on Architecture and Code Optimization},
  11(2):21:1--21:25, June 2014.

\bibitem{meantime-atc-2016}
A.~Farrell and H.~Hoffmann.
\newblock {MEANTIME: Achieving Both Minimal Energy and Timeliness with
  Approximate Computing}.
\newblock In {\em 2016 USENIX Annual Technical Conference (USENIX ATC 16)},
  pages 421--435, Denver, CO, June 2016. USENIX Association.

\bibitem{self-adaptive-software}
A.~Filieri, H.~Hoffmann, and M.~Maggio.
\newblock {Automated Design of Self-adaptive Software with Control-Theoretical
  Formal Guarantees}.
\newblock In {\em Proceedings of the 36\textsuperscript{th} International
  Conference on Software Engineering}, ICSE 2014, pages 299--310, New York, NY,
  USA, 2014. ACM.

\bibitem{gadiolo-2015}
D.~Gadioli, G.~Palermo, and C.~Silvano.
\newblock {Application Autotuning to Support Runtime Adaptivity in Multicore
  Architectures}.
\newblock In {\em {SAMOS XV}}, 2015.

\bibitem{approxhadoop-2015}
I.~Goiri, R.~Bianchini, S.~Nagarakatte, and T.~D. Nguyen.
\newblock {ApproxHadoop: Bringing Approximations to MapReduce Frameworks}.
\newblock In {\em Proceedings of the Twentieth International Conference on
  Architectural Support for Programming Languages and Operating Systems},
  ASPLOS '15, pages 383--397, New York, NY, USA, 2015. ACM.

\bibitem{hildebrandt-issta-2000}
R.~Hildebrandt and A.~Zeller.
\newblock {Simplifying Failure-Inducing Input}.
\newblock In {\em Proceedings of the 2000 ACM SIGSOFT International Symposium
  on Software Testing and Analysis}, ISSTA '00, pages 135--145, New York, NY,
  USA, 2000. ACM.

\bibitem{jouleguard-sosp-2015}
H.~Hoffmann.
\newblock {JouleGuard: Energy Guarantees for Approximate Applications}.
\newblock In {\em Proceedings of the 25\textsuperscript{th} Symposium on
  Operating Systems Principles}, SOSP '15, pages 198--214, New York, NY, USA,
  2015. ACM.

\bibitem{hoffmann-tpds-2012}
H.~Hoffmann, A.~Agarwal, and S.~Devadas.
\newblock {Selecting Spatiotemporal Patterns for Development of Parallel
  Applications}.
\newblock {\em IEEE Transactions on Parallel and Distributed Systems},
  23(10):1970--1982, Oct. 2012.

\bibitem{hoffmann-asplos-2011}
H.~Hoffmann, S.~Sidiroglou, M.~Carbin, S.~Misailovic, A.~Agarwal, and
  M.~Rinard.
\newblock {Dynamic Knobs for Responsive Power-Aware Computing}.
\newblock In {\em Proceedings of the Sixteenth International Conference on
  Architectural Support for Programming Languages and Operating Systems},
  ASPLOS XVI, pages 199--212, New York, NY, USA, 2011. ACM.

\bibitem{apex-2015}
K.~Huck, A.~Porterfield, N.~Chaimov, H.~Kaiser, A.~Malony, T.~Sterling, and
  R.~Fowler.
\newblock {An Autonomic Performance Environment for Exascale}.
\newblock {\em Supercomputing frontiers and innovations}, 2(3), 2015.

\bibitem{poet-2015}
C.~Imes, D.~H.~K. Kim, M.~Maggio, and H.~Hoffmann.
\newblock {POET: A Portable Approach to Minimizing Energy Under Soft Real-time
  Constraints}.
\newblock In {\em 21\textsuperscript{st} IEEE Real-Time and Embedded Technology
  and Applications Symposium}, pages 75--86, Apr. 2015.

\bibitem{stanford-snap}
J.~Leskovec.
\newblock {Stanford Large Network Dataset Collection(SNAP)}.
\newblock \url{http://snap.stanford.edu/data/}.

\bibitem{divya-2015}
D.~Mahajan, A.~Yazdanbakhsh, J.~Park, B.~Thwaites, and H.~Esmaeilzadeh.
\newblock {Prediction-Based Quality Control for Approximate Accelerators}.
\newblock In {\em Second Workshop on Approximate Computing Across the System
  Stack}, WACAS, 2015.

\bibitem{mithra-isca-2016}
D.~Mahajan, A.~Yazdanbaksh, J.~Park, B.~Thwaites, and H.~Esmaeilzadeh.
\newblock {Towards Statistical Guarantees in Controlling Quality Tradeoffs for
  Approximate Acceleration}.
\newblock In {\em 2016 ACM/IEEE 43\textsuperscript{rd} Annual International
  Symposium on Computer Architecture (ISCA)}, pages 66--77, June 2016.

\bibitem{openord-2011}
S.~Martin, W.~M. Brown, R.~Klavans, and K.~W. Boyack.
\newblock {OpenOrd: An Open-Source Toolbox for Large Graph Layout}.
\newblock volume 7868, 2011.

\bibitem{mierswa-2005}
I.~Mierswa and K.~Morik.
\newblock {Automatic Feature Extraction for Classifying Audio Data}.
\newblock {\em Machine Learning}, 58(2-3):127--149, Feb. 2005.

\bibitem{miguel-2014}
J.~S. Miguel, M.~Badr, and N.~E. Jerger.
\newblock {Load Value Approximation}.
\newblock In {\em Proceedings of the 47\textsuperscript{th} Annual IEEE/ACM
  International Symposium on Microarchitecture}, MICRO-47, pages 127--139,
  Washington, DC, USA, 2014. IEEE Computer Society.

\bibitem{misailovic-oopsla-2014}
S.~Misailovic, M.~Carbin, S.~Achour, Z.~Qi, and M.~C. Rinard.
\newblock {Chisel: Reliability- and Accuracy-Aware Optimization of Approximate
  Computational Kernels}.
\newblock In {\em Proceedings of the 2014 ACM International Conference on
  Object Oriented Programming Systems Languages \& Applications}, OOPSLA '14,
  pages 309--328, New York, NY, USA, 2014. ACM.

\bibitem{mitchell-1997}
T.~M. Mitchell.
\newblock {\em {Machine Learning}}.
\newblock McGraw-Hill, Inc., New York, NY, USA, first edition, 1997.

\bibitem{monteiro-2010}
S.~T. Monteiro.
\newblock {\em {Automatic Hyperspectral Data Analysis: A machine learning
  approach to high dimensional feature extraction}}.
\newblock VDM Verlag Dr. M{\"u}ller, 2010.

\bibitem{slambench-2015}
L.~Nardi, B.~Bodin, M.~Z. Zia, J.~Mawer, A.~Nisbet, P.~H.~J. Kelly, A.~J.
  Davison, M.~Luj\'an, M.~F.~P. O'Boyle, G.~Riley, N.~Topham, and S.~Furber.
\newblock {Introducing SLAMBench, a performance and accuracy benchmarking
  methodology for SLAM}.
\newblock In {\em IEEE International Conference on Robotics and Automation
  (ICRA)}, May 2015.
\newblock arXiv:1410.2167.

\bibitem{neapolitan}
R.~E. Neapolitan.
\newblock {\em {Learning Bayesian Networks}}.
\newblock Prentice Hall, 2003.

\bibitem{kinectfusion}
R.~A. Newcombe, S.~Izadi, O.~Hilliges, D.~Molyneaux, D.~Kim, A.~J. Davison,
  P.~Kohli, J.~Shotton, S.~Hodges, and A.~Fitzgibbon.
\newblock {KinectFusion: Real-Time Dense Surface Mapping and Tracking}.
\newblock In {\em IEEE ISMAR}. IEEE, Oct. 2011.

\bibitem{otaduy-sgp-2003}
M.~A. Otaduy and M.~C. Lin.
\newblock {CLODs: Dual Hierarchies for Multiresolution Collision Detection}.
\newblock In {\em Proceedings of the 2003 Eurographics/ACM SIGGRAPH Symposium
  on Geometry Processing}, SGP '03, pages 94--101, Aire-la-Ville, Switzerland,
  Switzerland, 2003. Eurographics Association.

\bibitem{palem-2005}
K.~V. Palem.
\newblock {Energy Aware Computing Through Probabilistic Switching: A Study of
  Limits}.
\newblock {\em IEEE Transactions on Computers}, 54(9):1123--1137, Sept. 2005.

\bibitem{quinlan-92}
J.~R. Quinlan.
\newblock {Learning With Continuous Classes}.
\newblock pages 343--348. World Scientific, 1992.

\bibitem{rinard-sc-2006}
M.~Rinard.
\newblock {Probabilistic Accuracy Bounds for Fault-Tolerant Computations That
  Discard Tasks}.
\newblock In {\em Proceedings of the 20\textsuperscript{th} Annual
  International Conference on Supercomputing}, ICS '06, pages 324--334, New
  York, NY, USA, 2006. ACM.

\bibitem{rinard-oopsla-2007}
M.~C. Rinard.
\newblock {Using Early Phase Termination To Eliminate Load Imbalances At
  Barrier Synchronization Points}.
\newblock In {\em Proceedings of the 22\textsuperscript{Nd} Annual ACM SIGPLAN
  Conference on Object-oriented Programming Systems and Applications}, OOPSLA
  '07, pages 369--386, New York, NY, USA, 2007. ACM.

\bibitem{ringenburg-asplos-2015}
M.~Ringenburg, A.~Sampson, I.~Ackerman, L.~Ceze, and D.~Grossman.
\newblock {Monitoring and Debugging the Quality of Results in Approximate
  Programs}.
\newblock In {\em Proceedings of the Twentieth International Conference on
  Architectural Support for Programming Languages and Operating Systems},
  ASPLOS '15, pages 399--411, New York, NY, USA, 2015. ACM.

\bibitem{rubio-gonzalez-sc-2013}
C.~Rubio-Gonz\'{a}lez, C.~Nguyen, H.~D. Nguyen, J.~Demmel, W.~Kahan, K.~Sen,
  D.~H. Bailey, C.~Iancu, and D.~Hough.
\newblock {Precimonious: Tuning Assistant for Floating-Point Precision}.
\newblock In {\em Proceedings of the International Conference on High
  Performance Computing, Networking, Storage and Analysis}, SC '13, pages
  27:1--27:12, New York, NY, USA, 2013. ACM.

\bibitem{samadi-2014}
M.~Samadi, D.~A. Jamshidi, J.~Lee, and S.~Mahlke.
\newblock {Paraprox: Pattern-Based Approximation for Data Parallel
  Applications}.
\newblock In {\em Proceedings of the 19\textsuperscript{th} International
  Conference on Architectural Support for Programming Languages and Operating
  Systems}, ASPLOS '14, pages 35--50, New York, NY, USA, 2014. ACM.

\bibitem{sage-2013}
M.~Samadi, J.~Lee, D.~A. Jamshidi, A.~Hormati, and S.~Mahlke.
\newblock {SAGE: Self-Tuning Approximation for Graphics Engines}.
\newblock In {\em Proceedings of the 46\textsuperscript{th} Annual IEEE/ACM
  International Symposium on Microarchitecture}, MICRO-46, pages 13--24, New
  York, NY, USA, 2013. ACM.

\bibitem{sampson-pldi-2011}
A.~Sampson, W.~Dietl, E.~Fortuna, D.~Gnanapragasam, L.~Ceze, and D.~Grossman.
\newblock {EnerJ: Approximate Data Types for Safe and General Low-Power
  Computation}.
\newblock In {\em Proceedings of the 32\textsuperscript{Nd} ACM SIGPLAN
  Conference on Programming Language Design and Implementation}, PLDI '11,
  pages 164--174, New York, NY, USA, 2011. ACM.

\bibitem{sampson-micro-2013}
A.~Sampson, J.~Nelson, K.~Strauss, and L.~Ceze.
\newblock {Approximate Storage in Solid-State Memories}.
\newblock In {\em Proceedings of the 46\textsuperscript{th} Annual IEEE/ACM
  International Symposium on Microarchitecture}, MICRO-46, pages 25--36, New
  York, NY, USA, 2013. ACM.

\bibitem{schkufza-pldi-2014}
E.~Schkufza, R.~Sharma, and A.~Aiken.
\newblock {Stochastic Optimization of Floating-Point Programs with Tunable
  Precision}.
\newblock In {\em Proceedings of the 35\textsuperscript{th} ACM SIGPLAN
  Conference on Programming Language Design and Implementation}, PLDI '14,
  pages 53--64, New York, NY, USA, 2014. ACM.

\bibitem{sherwood-asplos-2002}
T.~Sherwood, E.~Perelman, G.~Hamerly, and B.~Calder.
\newblock {Automatically Characterizing Large Scale Program Behavior}.
\newblock In {\em Proceedings of the 10\textsuperscript{th} International
  Conference on Architectural Support for Programming Languages and Operating
  Systems}, ASPLOS X, pages 45--57, New York, NY, USA, 2002. ACM.

\bibitem{sherwood-micro-2003}
T.~Sherwood, E.~Perelman, G.~Hamerly, S.~Sair, and B.~Calder.
\newblock {Discovering and Exploiting Program Phases}.
\newblock {\em IEEE Micro}, 23(6):84--93, Nov. 2003.

\bibitem{shoushtari-2015}
M.~Shoushtari, A.~BanaiyanMofrad, and N.~Dutt.
\newblock {Exploiting Partially-Forgetful Memories for Approximate Computing}.
\newblock {\em Embedded Systems Letters, IEEE}, Mar. 2015.

\bibitem{sidiroglou-douskos-fse-2011}
S.~Sidiroglou-Douskos, S.~Misailovic, H.~Hoffmann, and M.~Rinard.
\newblock {Managing Performance vs. Accuracy Trade-offs With Loop Perforation}.
\newblock In {\em Proceedings of the 19\textsuperscript{th} ACM SIGSOFT
  Symposium and the 13\textsuperscript{th} European Conference on Foundations
  of Software Engineering}, ESEC/FSE '11, pages 124--134, New York, NY, USA,
  2011. ACM.

\bibitem{capri-asplos-2016}
X.~Sui, A.~Lenharth, D.~S. Fussell, and K.~Pingali.
\newblock {Proactive Control of Approximate Programs}.
\newblock In {\em Proceedings of the Twenty-First International Conference on
  Architectural Support for Programming Languages and Operating Systems},
  ASPLOS '16, pages 607--621, New York, NY, USA, 2016. ACM.

\bibitem{reinforcement-learning}
R.~S. Sutton and A.~G. Barto.
\newblock {\em {Introduction to Reinforcement Learning}}.
\newblock MIT Press, Cambridge, MA, USA, first edition, 1998.

\bibitem{karthik-2015}
K.~Swaminathan, C.-C. Lin, A.~Vega, A.~Buyuktosunoglu, P.~Bose, and
  S.~Pankanti.
\newblock {A Case for Approximate Computing in Real-Time Mobile Cognition}.
\newblock In {\em Second Workshop on Approximate Computing Across the System
  Stack}, WACAS, 2015.

\bibitem{valiant-1984}
L.~G. Valiant.
\newblock {A Theory of the Learnable}.
\newblock {\em Communications of the ACM}, 27(11):1134--1142, Nov. 1984.

\bibitem{whang-icdm-2012}
J.~J. Whang, X.~Sui, and I.~S. Dhillon.
\newblock {Scalable and Memory-Efficient Clustering of Large-Scale Social
  Networks}.
\newblock In {\em Proceedings of the 2012 IEEE 12\textsuperscript{th}
  International Conference on Data Mining}, ICDM '12, pages 705--714,
  Washington, DC, USA, 2012. IEEE Computer Society.

\bibitem{seduction-2006}
D.~I. Wilson and B.~R. Young.
\newblock {The Seduction of Model Predictive Control}.
\newblock {\em Electrical \& Automation Technology}, pages 27--28, Dec. 2006.
\newblock {ISSN: 1177-2123}.

\bibitem{zafarani-liu-2009}
R.~Zafarani and H.~Liu.
\newblock {Social Computing Data Repository at {ASU}}, 2009.
\newblock \url{http://socialcomputing.asu.edu}.

\bibitem{zhu-popl-2012}
Z.~A. Zhu, S.~Misailovic, J.~A. Kelner, and M.~Rinard.
\newblock {Randomized Accuracy-Aware Program Transformations For Efficient
  Approximate Computations}.
\newblock In {\em Proceedings of the 39\textsuperscript{th} Annual ACM
  SIGPLAN-SIGACT Symposium on Principles of Programming Languages}, POPL '12,
  pages 441--454, New York, NY, USA, 2012. ACM.

\end{thebibliography}
}

\end{document}